# SalienTrack: providing salient information for semi-automated feedback in self-tracking with explainable AI

YUNLONG WANG, National University of Singapore

JIAYING LIU, University of Texas at Austin

HOMIN PARK, Institute for Infocomm Research

JORDAN SCHULTZ-MCARDLE, Chatham University

STEPHANIE ROSENTHAL, Carnegie Mellon University

JUDY KAY, University of Sydney

BRIAN Y. LIM, National University of Singapore

Self-tracking can improve people's awareness of their unhealthy behaviors and support reflection to inform behavior change. Increasingly, new technologies make tracking easier, leading to large amounts of tracked data. However, much of that information is not salient for reflection and self-awareness. To tackle this burden for reflection, we created the SalienTrack framework, which aims to 1) identify salient tracking events, 2) select the salient details of those events, 3) explain why they are informative, and 4) present the details as manually elicited or automatically shown feedback. We implemented SalienTrack in the context of nutrition tracking. To do this, we first conducted a field study to collect photo-based mobile food tracking over 1-5 weeks. We then report how we used this data to train an explainable-AI model of salience. Finally, we created interfaces to present salient information and conducted a formative user study to gain insights about how SalienTrack could be integrated into an interface for reflection. Our key contributions are the SalienTrack framework, a demonstration of its implementation for semi-automated feedback in an important and challenging self-tracking context and a discussion of the broader uses of the framework.

## 1 INTRODUCTION

Self-tracking and personal informatics have helped people to manage many areas of their lives, such as their finances [36,81], sustainability [6,40], physical activity [28,29], and diet [11,25,54,58,83]. Li et al. derived five stages of personal informatics and identified corresponding barriers [51]. Two critical stages are collection and reflection. Much research has focused on reducing the collection burden by automating data capture (e.g., with sensors [41,64], deep learning [18,54], and reinforcement learning [76]). Collection techniques now span manual, semi-automated, and automatic tracking [20]. In contrast, there has been little work on reducing the barriers to reflection. A few

Authors' Addresses: Yunlong Wang (yunlong.wang@nus.edu.sg) and Brian Y. Lim (brianlim@comp.nus.edu.sg), National University of Singapore, Singapore; Jiaying Liu (jiayingliu@utexas.edu), University of Texas at Austin, USA; Homin Park (park_homin@i2r.a-star.edu.sg), Institute for Infocomm Research, Singapore; Jordan Schultz-McArdle (j.schultzmcardle@chatham.edu), Chatham University, USA; Stephanie Rosenthal (srosenth@andrew.cmu.edu), Carnegie Mellon University, USA; Judy Kay (judy.kay@sydney.edu.au), University of Sydney, Australia.

exceptions include identifying effective visualizations [21,36], visualizing longitudinal data [22,84], speeding up information querying [47,60], providing auxiliary contextual data of tracked behavior [11], and using lightweight challenges [35]. This means people face challenges in making use of their copious tracking data.

We address this problem in the context of photo-based food logging, an important domain in a world where obesity poses huge health problems. Food logging apps have been shown to help users to gain awareness of their eating behaviors [30], identify unhealthy diet [11,63], improve nutrition intake [76], and control diet-related chronic diseases [24,65]. It is also a domain where people can readily a collect large amount of complex data covering their many eating episodes a day. This makes it challenging for people to reflect of the logged information.

We tackle this problem by creating a new approach, which we call SalienTrack. There are three key goals that drove our design. The first and central goal is to create a system which presents just the *salient* information. In the context of food tracking, this means that we need to identify the *salient events* (the subset of meals the user has photographed) and the *salient information* about them (aspects such as the nutrients, such as fat, and the cooking method, such as frying). A second core goal of our approach is to build an *explainable salience prediction model* — this means that an interface can enable a user to scrutinize the system's reasoning about salience of the information presented to them. The third goal for this work is to understand the right balance of user control in the interface for reflecting on salient event — essentially, there is a spectrum, with one extreme being completely *automatic* generation of information about a meal and the extreme opposite where users *manually* enter information for self-reflection [30]. Auto feedback would be less tedious for the user but may not support deep reflections. Hence, feedback mode should be selected to balance engagement and self-reflection.

We implemented SalienTrack with a machine learning and explainable AI-based approach to automatically provide concise, salient feedback based on the dimensions in the framework. To implement this, we pose two research questions: *RQ1) What information is salient in feedback? RQ2) How to automatically provide salient information in feedback?* We answer these questions in a two-phase pipeline shown in Figure 1. First, shown in the yellow box at the left, we conducted a field study of photo-based mobile food logging, where participants photographed their meals over 1-5 weeks, received Manual or Auto feedback at the end of each week, and rated the informativeness of the overall meal feedback and specific information types (e.g., macronutrients, cooking method). Next, shown in the orange box, we analyzed participant responses to identify useful features for feature engineering. Then, the green box shows that we trained a Gradient Boosted Tree prediction model with the field study data to build an explainable model to predict salient meals. To understand the model prediction and provide more salient feedback, the blue box shows how we used two explainable-AI techniques (SHAP [59] and Anchors [77]) to determine feature importance and counterfactual rules, respectively. These inform which feedback information (features) are most useful for self-reflection, and why. Finally, the violet box is for our qualitative user study on a set of prototype interfaces with different levels of automatic, manual and semi-automatic reflection interfaces. We make the following contributions:

1. Defined the SalienTrack Framework to identify dimensions for saliency.
2. Identified features to determine the informativeness and saliency in self-tracking feedback.
3. Implemented a prediction and explainable AI technique to select salient moments and information for self-tracking. We demonstrated it for mobile food logging with modeling and formative user studies.

Finally, we discuss how our approach can help to streamline self-tracking reflection for longer-term engagement, and how it can be generalized to other self-tracking activities.

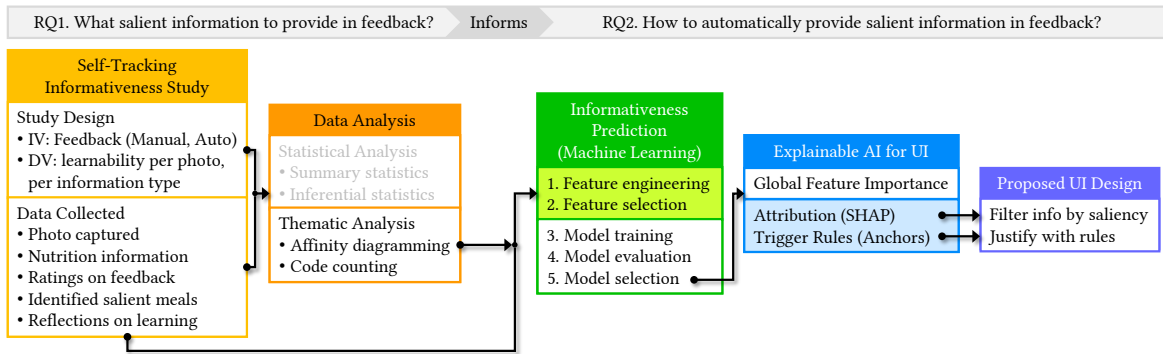

Figure 1: Project overview to answer two research questions of what self-trackers learn and how to provide the most salient feedback. We first conducted a field study, where participants tracked their diet and reviewed feedback. The results were analyzed with statistical (see Appendix 0) and thematic analysis. We then trained an informativeness prediction model on the collected data, and leveraged explainable AI techniques to develop more concise feedback to increase convenience and informativeness.

## 2 RELATED WORK

### 2.1 Self-Tracking for Food Journaling

The diversity of food and human eating behaviors complicates the self-tracking of diets and requires much information to be logged. From paper journals and questionnaires [5] to mobile logging using digital technologies and automated tracking using artificial intelligence, there are myriad methods for food tracking. Verbal and semantic information can be captured via highly scaffolded text-entry forms [2,30], or speech inputs [60]. However, these are burdensome for users over time. Conversely, the Ubicomp community has proposed many wearable sensing [85] approaches for more seamless tracking of eating behavior, such as with using wrist-worn [87], in-ear [9], or neck-worn devices [91]. However, these require custom hardware or atypical usage. In contrast, for this work, we leverage on the familiar practice of photographing meals with commodity smartphones. While merely capturing photos is reasonably good in aiding recall and reflection [30], advances in computer vision using deep learning have the potential to provide informative but less burdensome meal annotation. Several capabilities include automatically recognizing food dishes [68]; identifying food groups [79], ingredients [18] and cooking methods [17]; estimating calories [34], portion sizes [39,44], and drink healthiness [73]. This can be supplemented with other contextual information from smartphones, such as eating time [90] and location [63].

### 2.2 Self-Reflection from Feedback in Food Journaling

Although self-tracking need not be permanent [26], behavior change takes several weeks or months [82], so users need to be engaged for a moderate duration. We aim to sustain engagement by reducing the burden of reflection. Providing feedback frequently at each meal is very tedious and may dull the user's sensitivity towards the information [30]. Aggregating the feedback to once per day [30], once per week [11], or even longer periods [84] can reduce the frequency of review and facilitate deeper reflections. In this work, we chose the week duration to balance burden and reflection. More adaptive methods to reduce frequency include using AI to recommend the most appropriate moments for feedback based on preferences and contextual cues [53,76], These predict based on the outcomes of step count [53] and calories consumed [76]. However, these relate to behavior change outcomes which may be incidental or accidental [55]. Instead, we measured the perceived salience of each feedback and made predictions on them. Going beyond just increasing awareness, this considers how informative or useful the feedback is, rather than whether it was just noticeable. This corresponds to the early stages of *noticing* and *understanding* described by Kocielnik et al. [50] and the dimensions of *breakdown* and *inquiry* by Baumer [8].

Beyond just reducing feedback frequency to salient eating events, we aim to also select salient information about that meal to reduce information overload given the diverse information such as nutrition information (calories, ingredients, etc.) [11,47], context (e.g., events, places, and people [52]), and sensations [24,30]. Finally, some feedback interfaces require interactions and annotations, e.g., typing messages vs. multiple choice. Methods to reduce this burden include using various visualizations [15,21,36] or search-accelerators [47]. In this work, we explored providing feedback as automatically shown text or manually elicited data entry to balance between convenient passive learning and more engaging active learning [71].

## 3 SALIENT FEEDBACK IN PERSONAL INFORMATICS

### 3.1 Motivation and Application Use Case

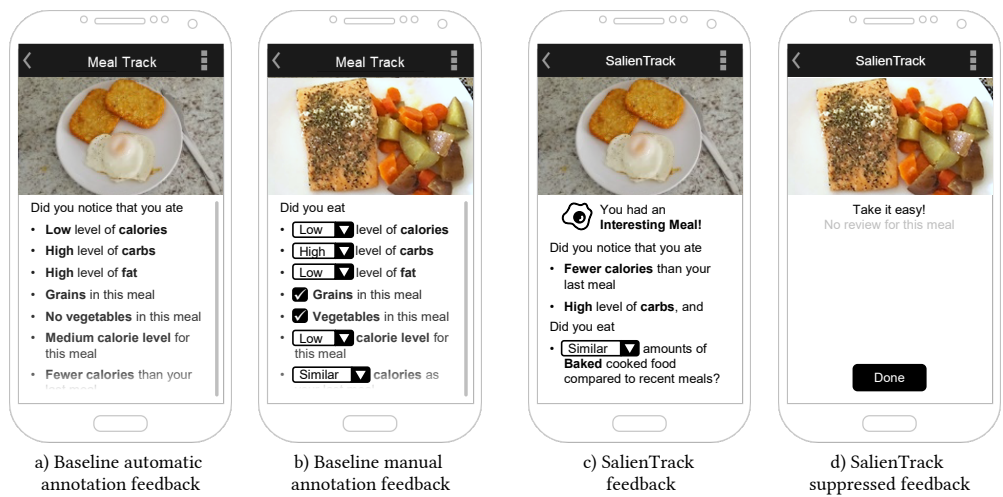

Figure 2: Examples of reflection on all (a,b) and salient (c,d) eating events. (a) baseline with detailed automatic (b) baseline detailed manual (c) automatically selected meal with just salient automated feedback with annotations (d) non-salient meal.

Figure 2 illustrates our motivation and approach for salient feedback. It shows two baselines, with Figure 2a presenting automated and comprehensive feedback about a meal. In the Figure 2b baseline, the interface requires users to manually enter the many details shown. Both baselines, which show all macronutrients and cooking information for all meals, is excessive and repetitive and may cause users to become disengaged. The manual one may well be better for supporting reflection, but it is also more tedious and not sustainable. Instead of either of these, we propose SalienTrack which aims to automatically select a salient subset of information to feedback to users. For example, healthy meals eaten by a typically healthy user or meals that are similar to recent ones could be omitted from feedback (Figure 2d) to be less patronizing or nagging. Furthermore, for meals selected for feedback, only more informative and salient aspects should be included to retain the user's limited attention (Figure 2c).

## 3.2 SalienTrack Framework

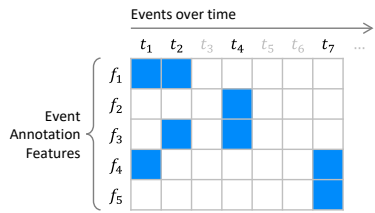

Figure 3: Schema of a tracked activity along two dimensions: events (columns) and annotations (rows). Each square denotes an annotation $f_\#$ for an event $t_\#$. Salient feedback should only include some events (e.g., $t_1$, $t_2$, $t_4$, $t_7$) and annotations (blue squares).

We define *salient* feedback in self-tracking as feedback containing the most important subset of information of a tracked activity that the user finds informative. As illustrated in Figure 3, we formulate a tracked activity as comprising multiple events over time, and each event as annotated with multiple features. Salient feedback would only select a subset of annotations (blue cells in Figure 3) instead of including all. Inspired by the Intelligibility framework by Lim and Dey [56], we introduce the SalienTrack Framework with question types to identify salient moments and annotations (Figure 4). First, we identify *when* salient moments that are most informative. Only some events are chosen for explicit feedback, while others are quietly logged. For self-tracking, these events can be exercise activities, meals, financial transactions, etc. Salient events can be selected heuristically [50], or with machine learning techniques, such as reinforcement learning [53,76] or supervised learning (our approach).

Second, we limit *which* information item to include in the feedback by identifying which features are more informative. Instead of providing a full description of the event (*inputs*) which can be overwhelming, this provides a concise subset of the most salient information. For example, the user could focus on the protein level for a particular meal, instead of all macronutrients. Salient factors can be obtained from user feedback (via surveys or focus group discussions), or data analysis (statistical analysis of significance, or explainable AI techniques). We employ SHAP [59] to identify influential attributions, and select the features with highest attribution for saliency. Note that this approach predicts what specific items to provide in feedback to promote informativeness, and not the annotation values for the tracked event (e.g., predicting nutrients from food). Third, we explain *why* the chosen features are informative. For example, a meal may be selected for feedback, because its fat content was >30g, which is high. These thresholds and rules can be obtained from domain expert specifications and literature, or through data mining methods (i.e., machine learning), which we employed with the Anchors [77] explanation method.

Finally, we determine *how* to provide the feedback for each salient feature. We explored two approaches: showing the information (auto-inform), or asking users to estimate the values (manual-elicit). For the latter manual-elicit approach, the application would not show feedback even if it has a prediction of the factor values. This ironic approach follows current practices for manual self-reflection, and can foster deeper reflection than auto-inform [8]. The choice between auto or manual can be made by the application designer or a scoring function by comparing the informativeness prediction confidence between both approaches.

We investigated this framework through stages in this work: 1) Dimensions of salient feedback (Section 3.2), 2) Measures of informative events and salient features (Section 4.2, Table 1), 3) Mechanisms for saliency selection (Section 5, Table 7), 4) Evidence to support the usefulness of the saliency dimensions (Section 7.1, Table 10).

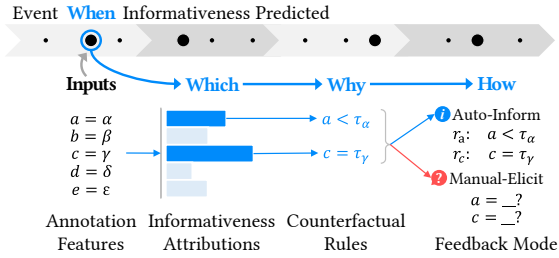

Figure 4: SalienTrack conceptual framework for salient feedback in self-tracking. This describes a chain of inquiry (blue text and arrows) for *when* to provide feedback, with *which* specific information items, with reasons *why* the user would learn from the event (as rules), and *how* to engage the user (with auto-generated information, or manual-elicited self-reflection). Given annotation feature values (inputs) of an event, predict the informativeness of its feedback; Latin alphabets represent variable names, Greek alphabets represent values and $\tau$'s represent rule thresholds. Informativeness Attributions (blue bar chart) indicate how important each feature is towards informativeness; most important features indicated with longer dark blue bars. Counterfactual rules explain why the event feedback is predicted to be informative; only rules of important features are included.

## 4 DATA COLLECTION: SELF-TRACKING INFORMATIVENESS STUDY

To support our goal to train a model to predict feedback informativeness and saliency, we need to collect training data and identify relevant features. We did this for the domain of food tracking. Specifically, we want to answer

*RQ1 What information is salient in feedback?*

Specifically, we aim to understand *when* users find feedback informative or not, *which* meal-specific information or annotations they find more informative, *why* the feedback was (not) informative, and how this differs across feedback modes. Prior works on mobile food journaling [30,54] focused on tracking meal details or providing feedback as an intervention or service. However, these did not explicitly measure the informativeness (or lack thereof) of the feedback in detail. Hence, we conducted a field study of mobile food tracking where participants logged their meals for at least one week (1-5 weeks), and reviewed their meals at the end of each week. We collected annotations for each meal and provided weekly feedback of all meals to situate users in the context of self-tracking, but focused on collecting data regarding the users' perceived informativeness of the feedback. This provides a labeled dataset of *when* feedback for a tracked event is informative, and informs *which* features could be useful for model training. We conducted the study with two feedback modes, Manual and Auto, to investigate and model *how* mode affects informativeness. Through data analysis and applying explainable AI, we will determine, for each meal, *which* information is salient and *why*. Next, we describe our method, apparatus, procedure, analyses and results.

### 4.1 Apparatus: User Interaction for Self-Tracking and Feedback

We designed a food tracking pseudo-app for two tasks — meal capture and weekly feedback. As participants logged their daily meals for several weeks, we conducted weekly surveys to provide meal feedback and report what they learned from the feedback. While much HCI and Ubicomp research focus on manual elicitation feedback due to their support for rich reflection [20,30,58,80], the burden on user review threatens their sustained use. Instead, much AI research [17,18,54,79] envision automatic feedback without user data entry. For generality, we conducted our data collection for Manual and Auto feedback modes. Among the different approaches for manual prompts (e.g., action plans [2], visual cues [36], meal enjoyment and context [30]), we chose to simply list nutrition information to align with basic food journals used by dietitians [47]. This also enables feasible automatic inference for Auto feedback.

*4.1.1 Photo-Based Food Tracking*

During the week, users photograph each meal; no annotation or data entry is needed. At the end of the week, users upload the images to our server to process the weekly feedback. This introduced a burden of requiring participants to remember to upload their photos, but this was manageable, since we successfully collected photos from many users. We leveraged existing applications rather than develop our own app to reduce development overhead, ensure app familiarity, reduce survey burden and fatigue, and improve app usage and study compliance [38,58].

*4.1.2 Weekly App Feedback*

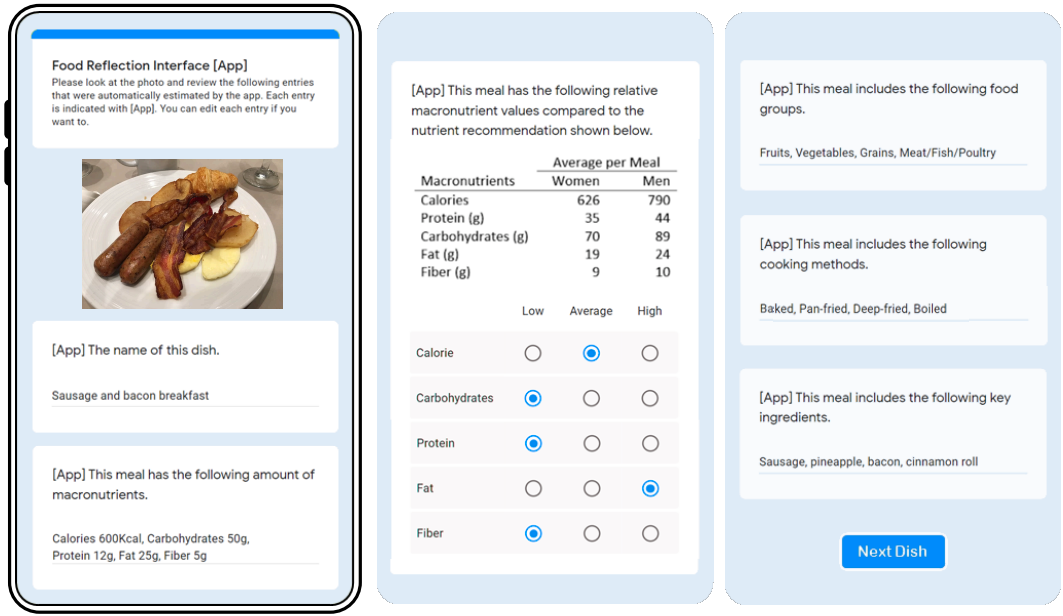

Figure 5: Example per-photo app feedback. Participants were presented with feedback information regarding the meal photo indicated with [App]. In Manual mode, all information is blank or unselected and users have to fill them out (see Figure 14 in Appendix); in Auto mode, all information is pre-filled and users can edit them (shown here).

Similar to [11], users received weekly feedback on their meals for the past week. We scheduled the feedback to be once per week rather than daily to: 1) allow more immersive reflection across multiple meals, 2) reduce reflection burden of reviewing feedback too frequently, and 3) enable feasible annotation for the Auto feedback. We implemented the app feedback with Google Forms, since it was sufficient to provide nutrition information and did not require maintaining our own database. Next, we describe the feedback information and interface (Figure 5).

The feedback comprises four types of nutrition information: Food Groups, Cooking Methods, Ingredients, and Macronutrients. We derived the nutrition feedback in close consultation with a trained dietitian. Macronutrients (calories, carbohydrates, protein, fat, fiber) are the most fundamental nutrition information, but are unintuitive for lay people to assess [7]. Thus, we include more explicit nutrition information. Food groups (fruits, vegetables, grains, meat/fish/poultry, and dairy) are the most intuitive information of food that people can easily perceive [30,45]. Ingredients provide more details about each meal. Cooking methods (baked, pan-fried, deep-fried, steamed, grilled, boiled, roasted) transform ingredients and affect their final calories and nutrients [16,46,86]. Providing these information types allows users to reflect at different granularity and depth. Though other information has been found to be useful (e.g., mood, post-meal satiety, social and physical contexts) [30,61], for feasibility, we only include information about the food dish that can be inferred from photo-based recognition and food databases.

All participants engaged with the same information, but had different interactions based on feedback mode. In Auto mode (Figure 5), users were shown the information to read or edit, e.g., how many calories were in the meal and what cooking methods were used. In Manual mode (Figure 14 in Appendix A.3), no information was provided and users answered questions to estimate the nutrition information, e.g., fat level and cooking methods in the meal. To reduce burden, we chose questions with multiple choice or short-text responses.

### 4.2 Measures for Feedback Informativeness, Perceived Accuracy and Burden

We focused on measuring the perceived informativeness of each meal feedback and specific information types to collect data to model salient feedback in terms of the four dimensions of the SalienTrack framework. Table 1 summarizes specific measures for each dimension. We also measure other secondary effects: perceived ease of understanding and tediousness (7-point Likert) when reflecting on the feedback, and perceived accuracy (5-point Likert) to assess whether the users are likely reflecting on wrong information.

Table 1: Measures in the data collection study to inform about the SalienTrack framework dimensions.

| Dimension | Measure |
| --- | --- |
| When | Per-meal: Informativeness rating (5-pt Likert scale: –2 to 2) of each meal feedback. See Appendix A.3 Q13. |
| Which | Post-week: Informativeness rating (7-pt Likert scale: –3 to 3) of each information type (e.g., macronutrients, food groups, cooking method) across all feedback. See Appendix A.4. |
| Why | Post-week: Text rationale for why specific meals and specific information types were informative. |
| How | Conducted data collection between-subjects for Manual and Auto feedback modes. |

### 4.3 Participant Recruitment and Study Procedure

We employed a remote recruitment and engagement approach to address several issues. First, the cuisine in our geographic location (non-western, non-United States) has limited nutrition data to prepare feedback. Second, the participants in our local culture are typically reticent. Hence, to widen our participant pool and align the participants' cuisine with online food nutrition information, we recruited US-based participants from Amazon Mechanical Turk (MTurk), and employed a remote engagement approach for longitudinal participation. This approach can also be used for conducting studies under social distancing requirements, such as during the Covid-19 pandemic [3,88]. Other benefits include higher participant diversity, and larger initial sample size to mitigate participant attrition. Similar methods for remote recruiting have been proposed for experiments with difficult recruiting requirements, such as field testing smart home technologies [12,13].

Figure 6 illustrates the participant recruitment process and study procedure. Participants were engaged through Amazon Mechanical Turk HITs for specific steps and incentivized to return since each step was paid, but may drop out at any time. Participants were compensated $0.05 USD, $0.70, $0.75, $8.00, for the screening, pre-survey, photo upload, and weekly survey HITs, respectively. The participant started with a screening survey testing instruction comprehension and basic nutrition literacy (Appendix A.1), which was reviewed within 2 days, and if passed, she was allocated to the Manual or Auto feedback mode, and invited to the pre-survey. The pre-survey (Appendix A.2) asked about the participant's demographics (age, gender, occupation, education, ethnicity, country of origin), attitudes towards healthy eating (i.e., self-assessment and motivation), weekly eating behavior (i.e., frequencies of eating specific food types and with cooking methods), and nutrition knowledge (adapted from [14]).

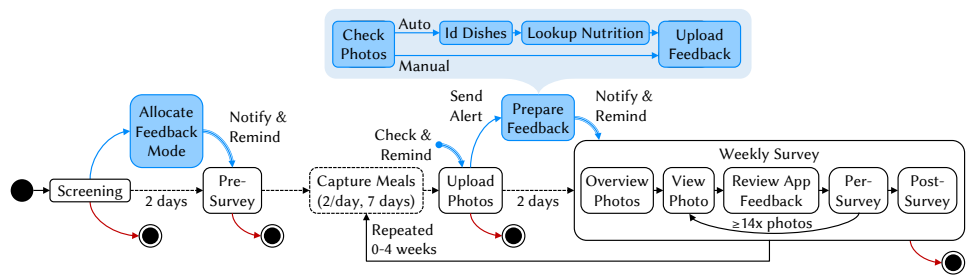

Figure 6: Field Study procedure showing participant interactions and study administrators' actions (blue) across week(s). Main stages involve recruitment, weekly data collection, and weekly surveys. Participants may drop out at various stages indicated by the concentric black circles, and are reminded to re-engage. Solid lines indicate immediate follow-up (black/blue is default, red is drop-out), dotted lines indicate follow-up after a few days, blue multi-stroke lines indicate repeated reminders (where needed).

The participant was instructed to photograph ≥2 meals/day, every day for 1 week. After 7 days, the participant uploaded her photos for feedback preparation. Research administrators checked the validity of photos and, for Auto, annotated the food names and nutrition information. Although the auto feedback was communicated as being done by a smart system, because of the accuracy limitation of current deep learning models, we implemented this with Wizard-of-Oz [43]. This is similar to using crowdsourcing annotators in PlateMate, which had good accuracy [72]. Future methods can use better image classifiers and automatic database look-ups. Food names, food groups, and ingredients were manually identified based on the annotators' knowledge and experience. Two annotators had extensive experience with western foods eaten in the United States. They trained other annotators and clarified when the latter were uncertain. Annotation was performed by looking up the food name in the MyFitnessPal food analysis database (https://www.myfitnesspal.com/food/search). Since the database contains potentially inaccurate crowdsourced data, annotators reviewed multiple entries and chose the 1st reasonable one. Finally, the annotations and photos were uploaded to a web server and the participant was notified to complete the weekly survey.

The weekly survey contained the food review feedback as an app and the post-survey. The participant verified her photos in an overview, then for each meal photo, she reviewed the app feedback (Figure 5 or Figure 14), and answered survey questions about perceived informativeness and perceived accuracy of the feedback (Appendix A.3). She then completed a post-survey (A.4 and A.5) about the week with questions to describe her most unhealthy and healthy meals, about her perceived informativeness of each nutrition information (overall, food groups, ingredients, cooking methods, and macronutrients), and about her user experience (ease of use and tediousness). Like in the pre-survey, she was asked about eating behavior and given the nutrition knowledge test to measure changes in knowledge. Finally, the participant indicated if she would like to continue another week or opt out.

### 4.4 Quantitative Results

We recruited participants from Amazon Mechanical Turk (AMT) with high qualification (≥5000 completed HITs with >97% approval) based in the United States (US). We screened 416 participants and invited 162 to the pre-survey. 136 participants completed the pre-survey; they were 70.6% female, and 39.0% ages 25-34, 32.4% ages 35-44, 16.2% ages 45-54. We expected poor commitment due to AMT tasks being primarily immediate and found that only 53 participants completed the first weekly survey. They were 34 female, 18 male, 1 nonbinary; 32.1% ages 25-34, 32.1% ages 35-44, 22.6% ages 45-54. Of these, 30 were allocated to the Manual feedback group, and 22 to the Auto group, though participation dropped off over time as expected (Table 2). In total, we collected 1,545 meal photos and per-meal survey responses, and 100 weekly survey entries.

Table 2: Number of participants in each group who completed the weekly survey for each week.

| Feedback Mode | Week 1 | Week 2 | Week 3 | Week 4 | Week 5 |
|---|---|---|---|---|---|
| Manual | 30 (100%) | 16 (53.3%) | 9 (30.0%) | 4 (13.3%) | 1 (3.3%) |
| Auto | 23 (100%) | 9 (39.1%) | 4 (17.4%) | 2 (8.7%) | 1 (4.3%) |

We present our findings on the primary measure of perceived informativeness, and secondary measures that supplement our understanding of the participant's experience. Our focus is on the user experience with meal feedback, and defer the supplementary analysis of background attitudes and food logging behavior in Appendix B.1 and B.2. To identify significant effects, we performed statistical analyses on user ratings detailed in Appendix B.3.

Participants found Manual feedback more tedious to use than Auto, especially in later weeks. They perceived the feedback as accurate (M=82.4% agreed). Not all feedback was informative (M=46.3%), suggesting the need to omit feedback sometimes. The perceived informativeness of Auto increased after the first week, but not for Manual. Further details are in Appendix B.4-B.6. Next, we describe the informativeness for specific information types.

We analyzed the relative differences in reported informativeness from different feedback information to inform which aspects are most salient. Participants appreciated learning more about diet behaviors than nutrition knowledge (Figure 7), but there was no difference across Feedback Modes. Among nutrition knowledge types, participants learned more about food groups and ingredients than cooking methods and macronutrients (Figure 7a). Among diet behavior information, participants learned more about the diversity of foods eaten than about whether they were eating more/less healthily (variation) (Figure 7b). These results highlight the need to include diet behavioral and temporal features to for salient feedback information. Among nutrition knowledge information, we also note that it is least useful to only inform about macronutrients, which many food logging apps typically do.

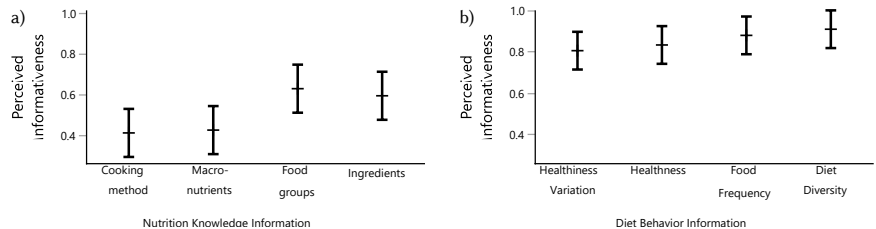

Figure 7: Results of weekly overall perceived informativeness for each Feedback Mode. Response values are binary (0 or 1); error bars indicate 90% confidence interval. Categorical axes sorted in ascending order of y-axis value.

### 4.5 Qualitative and Quantitative Analysis of Reflections about Perceived Informativeness

We performed a thematic analysis of participant rationales (n=692) in the weekly survey regarding what they found informative from the meal feedback. We used open coding [42] to derive categories and affinity diagramming [10] to consolidate categories to themes. Thematic coding was performed by one co-author researcher with regular discussion with co-authors. We then calculated inter-rater reliability on a random 15% subset of feedback coded independently by another co-author to obtain a Krippendorff's alpha with MASI distance [75] of $\alpha$=0.756, which indicates good agreement. We identified key themes in users' reflections based on what they learned from the meal feedback: cognitive space [50], valence of meal [11], contextual information, post-activity sensations [30], and agency [4,67]. These align with prior literature. We performed follow-up statistical analysis with linear mixed effects models to compare the difference of users' reflection between conditions. We discuss each theme next.

*4.5.1 Theme: Cognitive Spaces*

Participants reflected on three key types of information: nutrition knowledge, meal assessment, and diet behavior. This is similar to Kocielnik et al.'s description of a cognitive space with target and self domains [50]. *Nutrition knowledge* relates to just stating factual nutrition information and values about meals (macronutrients, food groups, cooking methods, ingredients); e.g., Participant $P_A11$ with Auto feedback learned that *"deep frying adds more calories than expected."* *Meal assessment* judges whether the meal or its nutritional component (e.g., calories) was healthy, above/below expectations, or how one should change the meal; e.g., $P_M17$ with Manual feedback assessed that her meal had *"so many carbs, no green veggies, and boxed fish."* *Diet behavior* describes a longer-term pattern across multiple meals by their averages, deviations (e.g., above/below, a lot), frequencies (e.g., often, never, seldom) and trends; e.g., $P_A22$ learned that *"there is much more variety in my dinners than in any other meal of the day."*

*4.5.2 Theme: Valence*

Participants often reflected on the feedback positively or negatively relative to healthiness. Blair et al. refer to this as the *valence* of the meal [11]. Positive and negative valence was similar across Feedback Modes, but varied across nutrition knowledge information (Figure 8). Participants discussed ingredients positively, typically by mentioning healthy ingredients, e.g., *"it was balanced with protein (egg) and fruit"* [$P_M5$], *"I had a good deal of veggies in this, as well as my organic grass-fed beef"* [$P_M6$]. Conversely, they discussed macronutrients negatively, perhaps due to common public health messaging about food healthiness; e.g., *"I learned that I eat a LOT of carbs …"* [$P_M18$], *"that there is less fiber in my food than i thought"* [$P_A13$]. These differences in valence have implications for diet feedback, such as prioritizing describing ingredients over macronutrients for positive, encouraging messaging [11,32,57].

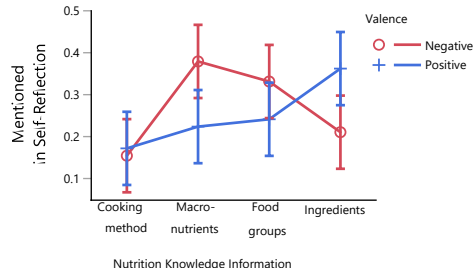

Figure 8: Self-reflection valence for different nutrition knowledge information. Responses are binary (0 or 1); error bars indicate 90%.

*4.5.3 Contextual Themes: Background, Agency and Control, and Post-Activity Sensations*

We identified other themes about the context surrounding the meal, namely, contextual background, agency and control, and post-activity sensations and feelings. Participants described the background *context* of their tracked meals to justify why a meal was healthy or unhealthy. E.g., $P_M17$ felt that "i*t was interesting to me that it was a relatively unhealthy week for me. It was around Father's Day so there were a few special meals peppered in (for my husband and my step-Dad), but my breakfasts were a bit heavier than normal so I should have had more cereal than the fatty stuff."* Conversely, $P_M8$ credited that *"the waitress suggested fruit and it actually was a really nice addition to my lunch."* Hence, participants sometimes attributed external factors for their diet choices. Additionally, many participants cooked their meals and proudly declared new cooking skills learned, e.g., $P_M12$ *"never grilled kabobs before so it was new to me."* This provided more justification for how they had or lacked <u>*agency*</u> to eat healthily. Finally, participants recorded how they felt when they ate certain meals (<u>*sensation*</u>). E.g., $P_M6$ reported that *"not* *having enough protein, fat and fiber made my body sluggish and I was rather cranky."* Some commented how certain ingredients were tasty, e.g., $P_A12$ learned that *"pineapple tastes good when grilled"*; or how certain ingredients were satiating, e.g., $P_M8$ *"used a tortilla instead of the hash browns. I was surprised that it kept me feeling fuller for longer, because of the fiber."* These suggest their awareness of new incentives to eat such dishes more in future.

Similar to [74] that found that photo-taking (i.e., tracking) behaviors differed with annotation automation, we found that users reflected differently based on whether the feedback was automatically shown or manually elicited. These reflections were mentioned at different frequencies across feedback modes. Contextual information was mostly written by participants with Manual feedback and almost never mentioned by participants with Auto feedback (only 9 mentions). Participants reflected most about their agency (or lack thereof), followed by contextual background, and sensation after the meal (Figure 9). Participants with Manual feedback reflected more about Agency (M=.079 vs. .010, p<.0001) and Context (M=.056 vs. .010, p=.0020) than participants with Auto feedback.

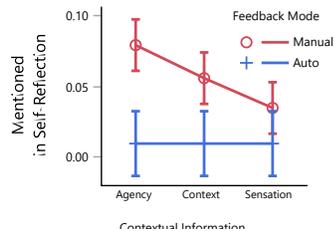

Figure 9: Self-reflection mentions about contextual information. Responses are binary (0 or 1); error bars indicate 90%.

### 4.6 Summary of Findings and Implications for Design

In summary, answering RQ1 about what information is salient in feedback, we found that participants:
1. Perceived the meal feedback as accurate and found Auto feedback easier to use than Manual feedback.
2. Reported learning more about their diet behaviors than nutrition knowledge.
3. Reflected along the target-self cognitive space: nutrition knowledge, meal assessment, diet behaviors.
4. Reflected on different nutrition information type depending on positive or negative valence.
5. Reflected more about contextual information with Manual feedback than with Auto feedback.

These findings pose some implications for design, namely,
- Scaffold feedback with cognitive spaces (target domain, self-assessment and long-term behavior awareness).
- Provide Auto and Manual feedback together for nutrition-specific informativeness and contextual reflection.
- Prioritize feedback for each nutrition information type to support positive or negative valence in reflection.

### 5 SALIENTRACK MODEL PREDICTION AND EXPLANATION FOR CONCISE FEEDBACK

With the data and findings from the data collection study, we propose a technology and technique to provide salient feedback to balance reflection burden and informativeness. This answers our second research question:

*RQ2: How to automatically provide salient information in feedback?*

This involves three technical steps (Figure 1, right): informativeness prediction, explainable AI for the user interface (UI), and proposed UI design. The machine learning approach has three parts (Figure 10) for annotation prediction, informativeness prediction, and informativeness explanation. The first part is to train a convolutional neural network $M_y$ for automatic recognition of the food from a meal photo $x$, including predicting or looking up nutrition information. We denote these nutrition annotations as $\breve{y}$. Modeling for this is well-established [54,68] and we defer providing further detail. Note that SalienTrack is premised on automatically recognizing meal photos, though feedback may be conveyed as manual elicitation or automatic display. Second, we propose to predict an informativeness score $\hat{\imath}$ by training model $M_l$ on diet features $\breve{f}$ based on aggregate and temporal variables extracted from the nutrition annotation $\breve{y}$, with heuristic preprocessing method $\mathcal{M}_f$. This allows nutrition knowledge and diet behavior information to be encoded. The informativeness model $M_l$ will predict whether a user is likely to learn much from each specific photographed meal. If the informativeness score is high, then feedback should be provided for that meal, otherwise, feedback should be omitted. Third, for concise feedback, we use model explanations to only show salient features and rules. Our approach to exploit explanations differs from typical uses of explainable AI. We are not proposing to explain the primary prediction task, i.e., why $M_y$ predicted the dish name or calorie level. Instead, we use explanations from $M_l$ that explains how diet features $\breve{f}$ influenced the current informativeness score $\hat{\imath}$. The explanations will first determine importance weights $\widehat{w}$ for all annotation features, and rules $\hat{r}$ for some features, then filter the rules to only the most salient ones $\hat{r}_w$.

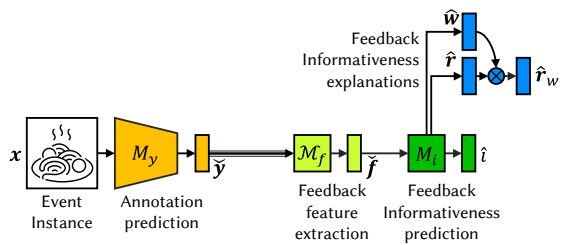

Figure 10: Overview of proposed SalienTrack system with modules for annotation prediction (yellow), informativeness prediction (green) from multi-meal diet features (yellow-green) and informativeness explanations (blue). We propose informativeness prediction as a key capability for providing salient feedback in self-tracking. The multi-line arrow indicates aggregate features extracted across multiple meals. The cross symbol refers to a masking operation with importance weights $\widehat{w}$ on the rules $\hat{r}$.

### 5.1 Diet Feature Extraction

Our quantitative and qualitative analyses of the users' review of food logging feedback identified different information and aspects that users learned and reflected on. We aimed to extract data features spanning different cognitive spaces on *nutrition knowledge*, *meal assessment*, and *diet behavior* from meal annotation and survey data from our earlier study. The premise is that nutrition knowledge and meal assessment annotations can be automatically inferred with image-based classifiers [54,68], though for our initial study, we depend on manual annotation by our previous participants and research annotators. We added features for the number of food groups and ingredients to capture the *diversity* in the meal. Diet behavior features calculated from the historical meal records include the mean, standard deviation, trend (slope of linear interpolation), change from previous meals, maximum of each nutrition information across different time periods (recent 2-4 meals, recent 2 meals with the same meal type). These features are consistent with meal annotations, participant reflections, and literature on dietetics [62]. We excluded features about user demographics (age, gender) and study treatment (week) so that the models trained are generalizable beyond the previous user study. Altogether, this produces 580 data features, which can suffer from the curse of dimensionality. To reduce dimensionality, we selected features using recursive feature elimination [41] for tree-based models, and mutual information-based univariate feature selection method [71] for other models. The final model was trained with 30 selected features including food habits, nutrition knowledge, and diet behaviors, as shown in Table 12 in Appendix C.1.

## 5.2 Informativeness Prediction

Table 3: Performance evaluation of various models trained on manual (left) and auto (right) feedback datasets. First two models are intrinsically interpretable but less accurate than the bottom three. Best performance results bolded.

| Model | Manual Feedback Dataset | | | Auto Feedback Dataset | | |
|---|---|---|---|---|---|---|
| | Accuracy | F1 score | PR AUC | Accuracy | F1 score | PR AUC |
| Logistic Regression | 0.70 | 0.66 | 0.68 | 0.68 | 0.77 | 0.81 |
| Decision Tree | 0.73 | 0.71 | 0.65 | 0.68 | 0.74 | 0.73 |
| Multi-Layer Perceptron | 0.72 | 0.67 | 0.74 | 0.73 | 0.80 | 0.84 |
| Random Forest | 0.76 | 0.71 | 0.78 | **0.76** | 0.78 | 0.83 |
| XGBoost | **0.80** | **0.74** | **0.80** | **0.76** | **0.84** | **0.89** |

We modeled informativeness prediction as a classification problem by binarizing the self-rated informativeness (rating >0 or not). Our dataset had 1,545 instances (922 and 623 from Manual and Auto, respectively). The dataset is balanced with 46.3% of instances labeled with high informativeness. We investigated 5 machine learning models. Logistic regression and decision tree were considered for interpretability, but they sacrifice model performance. Multi-layer perceptron (neural network), Random Forest, and Gradient-Boosted Trees (XGBoost) [19], were considered for accuracy, but are less interpretable due to their large number of model parameters.

Informed by different reflection behaviors of participants in the data collection study, we trained models separately on the Manual and Auto datasets to understand how features variously influence the informativeness for each feedback mode. Models were evaluated with 5-fold cross-validation. Table 3 summarizes the model evaluations, reporting various metrics to compare models. The interpretable models had poorer performance than the larger models. We selected XGBoost for SalienTrack, since it had the highest performance with F1 scores 0.74 and 0.84 for Manual and Auto, respectively. This indicates good prediction performance for binary classification.

## 5.3 Saliency and Insights from Informativeness Explanations

We investigated model explanations to 1) understand how the model made decisions regarding high or low informativeness, and 2) use as a mechanism to provide salient feedback to end-users. We employed SHAP and Anchors explanations to see which features were important and why they affected informativeness, respectively. We describe how they are calculated and interpreted, and evaluate their correctness towards saliency.

### 5.3.1 Instance Explanation

SHAP [59] calculates the attribution by each feature towards the model's inference for a specific instance. For each instance prediction, the attributions inform how important each feature is (magnitude) and whether it influences the decision towards informativeness (positive sign or large) or not (negative or small). Consider the example in Table 4 about a meal that $P_M8$ rated as informative (Rating=2 on Likert scale –3 to 3). SalienTrack had predicted high informativeness for both Manual and Auto feedback models. However, both models had slightly different explanations. For Auto feedback, features "Meal Macros (Calorie level) : Change[Prev1-Current]", "Meal Macros (Carbs level)", and "Meal Cooking (Pan/Air Fried) : Mean[Prev3-Current]" were the most influential towards high informativeness, while "Meal Macros (Calorie level)" and "Meal Macros (Fat level)" suggested low informativeness. For Manual feedback, "Meal Macros (Fat level)" about the current meal, and "Meal Cooking (Baked) : SD[Prev2-Current]" and "Meal Macros (Calorie level) : Highest[Prev3-Current]" about recent meals were the most influential towards high informativeness, while "Meal Macros (Protein level)" and "Meal Food Groups (Vegetables)" suggested low informativeness. Together, these suggest that the user could learn more from being automatically shown the decreased calorie level of her meal, and being asked to manually indicate if recent meals typically had baked cooking. From the SHAP attributions, we select the top attributing features towards informativeness (darker color), and exclude features about prior habits, since they are less actionable. These SHAP attributions change for different instances (see global visualization in Appendix C.3 Figure 19). Thus, saliency is dynamic with feedback instances.

Table 4: Explanations of a meal of "Fried egg and hash browns" predicted as high informativeness for both Manual and Auto Feedback models. Feature attribution (SHAP) is shown as vertical bar charts with lowest attribution set to 0. Grey bars are for user profile features, red for informativeness towards Manual feedback and blue for Auto feedback. Darker colored bars indicate the most salient features. Here, Anchor rules explain why the meal feedback was predicted as highly informative.

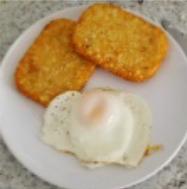

| | Feature Name | Feature Value | Manual Feedback Model SHAP Attribution | Manual Feedback Model Anchor Rule | Auto Feedback Model SHAP Attribution | Auto Feedback Model Anchor Rule |
|---|---|---|---|---|---|---|
| | Prior Eating Habit (Vegetables) | Eats 4-6x/week | | | | |
| | Prior Eating Habit (Fruit) | Eats 1-3x/week | | | | |
| | Meal Macros (Calorie level) | Low | | | | |
| | Meal Macros (Carbs level) | High | | | | ≥ High |
| | Meal Macros (Protein level) | Low | | | | |
| | Meal Macros (Fat level) | High | | ≥ High | | |
| | Meal Macros (Fiber level) | Low | | | | |
| Fried egg, and hash browns | Meal Food Group (Grains) | Has | | | | |
| | Meal Food Group (Vegetables) | None | | | | |
| | Meal Food Group (Meat) | None | | | | |
| | Meal Food Group (Fruits) | None | | | | |
| | Meal Food Group (Dairy) | Has | | | | |
| | Meal Food Groups Count | 2 | | | | |
| | Meal Cooking (Baked) | None | | | | |
| | Meal Macros (Calorie level) : Mean[Prev1-Current] | Medium | | | | |
| | Meal Macros (Calorie level) : Highest[Prev3-Current] | High | | > Low | | |
| | Meal Macros (Protein level) : Highest[Prev3-Current] | High | | | | |
| | Meal Macros (Fat level) : Highest[Prev3-Current] | High | | | | |
| | Meal Macros (Fat level) : Change[Prev2-Current] | Unchanged | | | | |
| | Meal Macros (Calorie level) : Change[Prev1-Current] | Decreased | | | | ≤ Decreased |
| | Meal Food Group (Vegetables) : Change[Prev2-Current] | Decreased | | | | |
| | Meal Food Group (Vegetables) : Change[PreSameMealType-Current] | 0/2 | | | | |
| | Meal Ingredients Count : Highest[Prev2-Current] | 4 | | | | |
| | Meal Cooking (Microwaved) : Mean[Prev1-Current] | 0/2 | | | | |
| | Meal Cooking (Microwaved) : Mean[Prev3-Current] | 0/4 | | | | |
| | Meal Cooking (Pan/Air Fried) : Mean[Prev3-Current] | 3/4 | | | | > 2/4 |
| | Meal Cooking (Baked) : SD[Prev2-Current] | Medium | | > Low | | |
| | Meal Cooking (Deep Fried) : SD[Prev2-Current] | Medium | | | | |
| | Meal Cooking (Raw) : SD[Prev3-Current] | Low | | | | |
| | Meal Cooking (Steamed) : Trend[Prev3-Current] | Unchanged | | | | |

An Anchor explanation [77] finds a set of counterfactual rules for the instance that when not satisfied causes the model to change its inference. In the example shown in Table 4, for Manual feedback, the rule "Meal Macros (Fat level) ≥ High" indicates that had the fat level not been high, the feedback would not be informative. For Auto feedback, the rule "Meal Change (Calorie level) : Change[Prev1-Curent] ≤ Decreased" indicates that if the current meal had not decreased in calorie level from the previous 1 meal, the model would predict low informativeness instead. Table 13 (Appendix C.2) shows SHAP and Anchor explanations of a meal with low informativeness.

*5.3.2 Evaluating Correctness of Saliency Selection*

We evaluated explanation correctness to determine whether the most salient features most affected predictions. For an instance prediction, a feature that is more important will cause the prediction confidence to change more if the feature value is changed. We induce the change by perturbing salient features across the counterfactual rule threshold, e.g., for the instance in Table 4 predicted as informative, for the feature "Meal Cooking (Pan/Air Fried) : Mean[Prev-Current]" with value 3/4 and Anchor Rule explanation >2/4, we would change its value to 2/4 to just violate this rule. Specifically, we create a counterfactual instance with only that feature value change, have the model predict the instance's informativeness, and measure the decrease in prediction confidence of informativeness. For an informative/uninformative prediction, we expect that this perturbation should decrease/increase the confidence to indicate the correct influence. Mathematically, we calculate the Signed Prediction Confidence Change for the $k$th feature $x_k$ as $\Delta p_k = -y(p - p_{\neg r_k})$, where $y$ is 1 if the prediction is positive and –1 if the prediction is uninformative, $p$ is the original prediction informativeness for the instance, and $p_{\neg r_k}$ is the prediction confidence for the counterfactual instance. Figure 11 validates that, on average, features that were ranked as more important were more influential in determining whether the instance is predicted as informative or not.

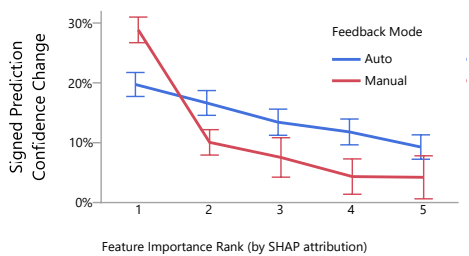

Figure 11: Evaluation of saliency correctness for the top 5 most important features for instances in a test dataset. Higher Signed Confidence Change indicates that the feature is more important.

## 5.4 Semi-Auto Feedback with Dynamic Selection of Manual and Auto Modes

Having trained two models for saliency prediction for Manual and Auto feedback, we further propose a third variant Semi-Auto that combines both feedback modes. This will limit the burden of always requiring manual annotation by more often providing automatic feedback and occasionally providing manual feedback. Providing feedback with both Manual and Auto feedback modes involves several steps. First, we calibrate the relative desire for Manual or Auto feedback by using preference weight $\alpha^F$, for each feedback mode $F \in \{\text{Manual}, \text{Auto}\}$. This depends on the application designer and user, e.g., $\alpha^{\text{Auto}}$ should be higher to prioritize lower burden. Then we select the top $k$ features with highest SHAP attributions from the maximum across feedback modes, i.e., $\widehat{w}_k = \max(\alpha^{\text{Manual}} \widehat{w}_k^{\text{Manual}}, \alpha^{\text{Auto}} \widehat{w}_k^{\text{Auto}})$. We chose $k = 3$. Each top feature is conveyed with its maximizing feedback mode, i.e., $F_k = \text{argmax}_F \alpha^F \widehat{w}_k^F$. Hence, for a tracked event selected for feedback, the three features that are selected for feedback are either shown automatically for plain reading, or require the user to manually estimate its value.

## 5.5 Summary of Model Predictions and Explanations for Salient Feedback

We have developed the salient feedback model with an informativeness model prediction and explanation techniques to select salient instances, features, and rules for different feedback modes. Table 5 summarizes how the mechanisms implement the dimensions in the SalienTrack framework (Figure 4).

Table 5: Mechanisms in the explainable prediction model for dimensions in the SalienTrack framework.

| Dimension | Mechanism |
|---|---|
| When | Treat events as instances with annotations as features, and predict informativeness as binary classification. |
| Which | Calculate SHAP attributions for the instance features, rank order features by attribution magnitude towards the prediction, and filter saliently select (filter) the top $k$ features ($k$ chosen by the application designer). |
| Why | Calculate Anchor rules to explain the criteria for the instance being predicted as informative or not. |
| How | Compare prediction confidences for the Manual and Auto models and select mode with higher confidence. |

## 6  FORMATIVE STUDY OF SALIENTRACK FEEDBACK

We formatively studied the usefulness of the SalienTrack feedback interface with a scenario-driven semi-structured interview study. This allowed us to verify positive aspects of SalienTrack and identify issues before further investment in engineering and field testing. We aimed to qualitatively examine the prospective usefulness of dynamically selecting fewer features in feedback, understanding whether users would prefer other information, and how explore their opinions regarding feedback mode (manual, auto, semi-auto).

### 6.1  Experiment Apparatus: Baseline and SalienTrack Feedback User Interface

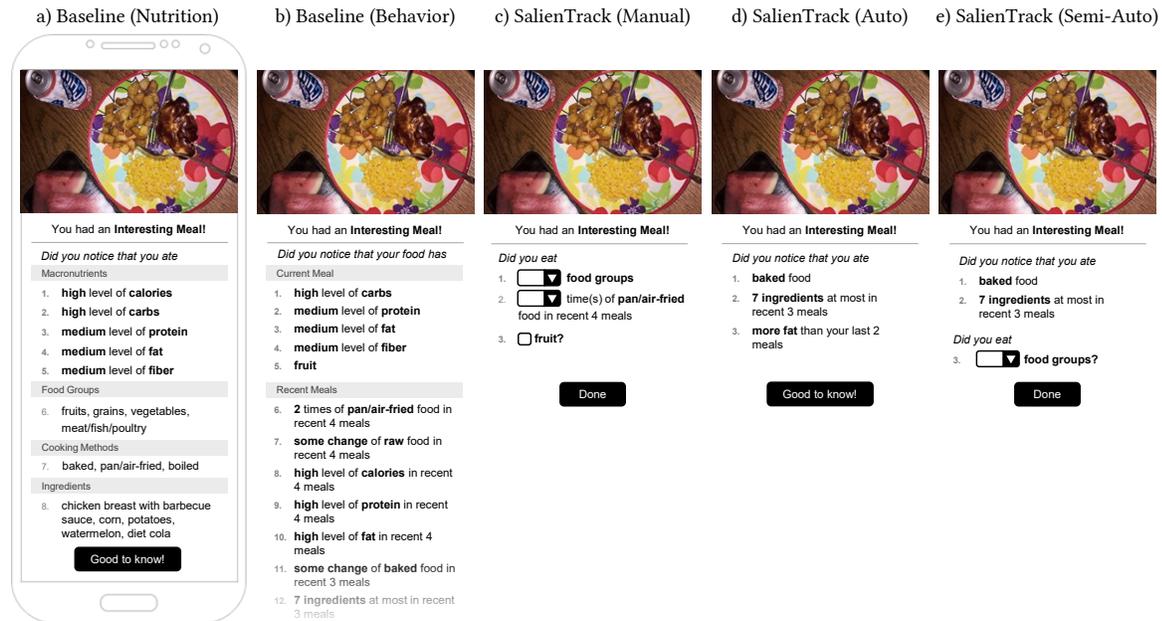

Figure 12: Example UI mockups of two Baseline and three SalienTrack variants used to probe opinions in the formative study.

We were interested to compare the features in SalienTrack and varied Feedback Modes as the independent variable with two baseline conditions (Baseline-Nutrition, Baseline-Historical), and three SalienTrack conditions (Manual, Auto, Semi-Auto). See for Figure 12 example screenshots. We prototyped static mock-ups of app screenshots in PowerPoint slides rather than an interactive prototype. This is equivalent to interviewing with paper prototypes to elicit more open discussions from the participant, since the interfaces look less developed. All feedback interfaces share the same basic interface design comprising a list of features with each feature value either automatically shown (e.g., "Low level of calories"), or requiring manual entry in a simple form. For long lists in the baseline interface, we grouped related features by categories (e.g., macronutrients, food groups, recent meals) to aid interpretability. For Auto feedback, users only need to read the values and do not need to estimate them. The Manual entry interface uses dropdown menus or checkboxes to limit choice overload and reduce user burden. More appealing and sophisticated visualizations could be used in future work (e.g., [36,84]) but our focus was on the straightforward truncation of features selected for saliency. The feedback can also incorporate *why* the meal was salient by displaying rules with numeric quantities; for categorical values, rules are already implicit by just showing the selected value (equality). For simplicity in the formative study, we used categorical features.

We examined three variants of SalienTrack with the Manual, Auto, and Semi-Auto modes. We limited to the top three salient features to limit information overload. Furthermore, if SalienTrack predicts low informativeness, then feedback would be omitted from the set of meals to review. Since SalienTrack selects features based on nutrition information and historical behavioral characteristics (e.g., average in past 3 meals), we had two baseline conditions to investigate their base usefulness; unlike SalienTrack, features for these baselines were curated and fixed selections for each meal, not dynamically selected. Baseline-Nutrition feedback shows 8 items in four categories (macronutrients, food group, cooking methods, ingredients). Baseline-Behavioral feedback shows the top 15 items selected based on feature importance of the XGBoost model; we divided these into two categories (current, recent meals). The long list for Baseline-Behavioral feedback also allows us to investigate the impact of lengthy feedback which may lead to information overload. All baseline feedback were in Auto mode, since requiring their manual entry would be obviously tedious. Although random subset selection can be considered a baseline to compare against SalienTrack selection, we did not include this since it would be perceived as clearly arbitrary and less useful.

### 6.2 Experiment Procedure

The experiment procedure was as follows. We first briefed the participant about the scenario of photographing multiple meals over time and having a review session of several images through a feedback app. The participant was then instructed on how each of the five Feedback Modes worked and what information they provided. Details for the briefing and tutorial are in Appendix D. After the briefings, we commenced the main experiment. We had selected new food meal images that were canonical of our training dataset (western dishes typical of our data collection), and generated baseline and SalienTrack feedback. Participants were instructed to imagine being in a scenario where the user has eaten 7 consecutive meals and was reviewing the last 4 meals using a feedback app.

In the main study, first, the participant chose 1 set of meals from 3 possible sets that she was most familiar with to analyze. This was to maximize the familiarity and relevance of the meals to the participant's diet, and mitigate issues when interviewing on scenario data. The participant viewed 4 trials in the main study. In the first trial, the participant was shown three prior meals eaten (only as photos) to contextualize the scenario, and the app feedback for a fourth meal. The subsequent three trials showed the next meal in sequence and incremented the recent meals by one as a sliding window. The feedback was shown for all 5 Feedback Modes in the order: Baseline-Nutrition, Baseline-Behavior, SalienTrack Manual, Auto, Semi-Auto. We showed food images and app screenshots in a PowerPoint presentation, one screenshot at a time, and asked the participant to describe which information she found useful, that she could learn from, or what other information she would prefer to learn, and what she found tedious. We provided clarification when questions were raised. After reviewing four meals across all the trials, we asked the participant to rank and explain the informativeness and tediousness of each Feedback Mode, and discuss any features she would like to have included or excluded.

## 6.3 Experiment Findings

We recruited 10 participants through convenience sampling from people residing in the US, since they would be more familiar with US-based foods of our dataset. They were 3 male and 7 female, ages 26 to 35. We interviewed participants over Zoom and recorded the audio and screen interactions for subsequent analysis. Each interview session lasted about one hour, and each participant was compensated with a USD $15 Amazon gift card.

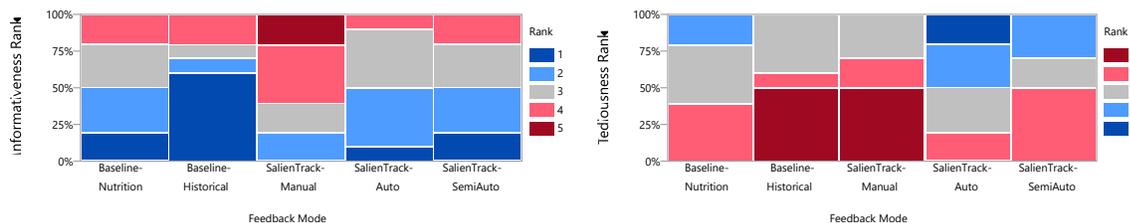

Figure 13: Participant ranking of baseline and SalienTrack feedback modes for Informativeness and Tediousness.

Participants ranked the five Feedback Modes by informativeness and tediousness (Figure 11). As expected, Manual feedback was most tedious and least informative, since users needed to know and enter the information themselves. Participants most often ranked Baseline-Historical feedback as the most informative, but also most often ranked it as most tedious, because of its long list. Participants found Baseline-Historical information more useful than Baseline-Nutrition, because the latter only had information about the current meal and not previous ones. This shows that providing excessive feedback details for each meal is less appreciated than showing more meals. Participants found SalienTrack-Auto the least tedious as expected, but found SalienTrack-SemiAuto more informative due to the complementary benefits of mixing both Auto and Manual feedback. The key take-away is that SalienTrack-SemiAuto balances between reducing tediousness and improving informativeness. Next, we qualitatively analyzed the explanations by participants for their opinions.

We performed a thematic analysis of participants' utterances using open coding [42] guided by the SalienTrack Framework dimensions, themes from our data collection study, and objectives to examine usefulness across feedback modes. Thematic coding was performed by one co-author researcher with regular discussion a senior co-author. Participants liked the dynamic selection and concise feedback, and the variety of feedback modes, though they wanted more information on demand, and feedback in context of their goals. We report the most salient details.

**Focused learning on nutrition information and diet behavior (Baseline-Nutrition vs. Baseline-History).** Participants reflected on the feedback based on cognitive space similarly to our previous participants in the data collection study (Section 4.5). Some participants focused on specific nutrition information instead of the full long list with many information types. P7 said, *"I only care about the macronutrients, especially the calories and carbs, because I track my food mostly for losing weight."* Similarly, P3 mentioned that *"I would check the food groups in the current meal and recent meals to see if my meals were balanced."* Conversely, some participants were also interested in diet behavior. P4 felt that *"the average level of macronutrients is very helpful for me. Actually, I also want to know the calorie level of the whole week."* Therefore, the categorical grouping provided in the baseline feedback was useful.

**Preference for concise but extensible feedback (Baseline vs. SalienTrack).** Most participants appreciated the concise, dynamic information in SalienTrack compared to the longer lists in the baselines. P7 most preferred SalienTrack-SemiAuto because she thought that *"I literally have to do it [food tracking] every day, so I want to choose the easiest one, you know, the one that I think it's more interesting and more interactive. That's why I choose Semi-*

*Auto because, you know, they highlight some of the facts, and then I get to answer the questions or give them estimates."* P4 mentioned that *"it's hard to pay attention to this kind of detailed information [in Baseline-Historical] for a long time to get valuable information."* P1 liked the dynamic nature of SalienTrack and mentioned that *"it's good the AI select different information for each meal. Overall, it will cover a wide range of information types"*. Interestingly, some participants wanted to request for more information. P2 said *"the three pieces of information draw my attention, but I want to explore more information. Like this one, the average fat level in recent three meals is low, but what about the carbs and protein?"* P9 commented that *"two of the information [in SalienTrack-Manual] are interesting but the third one is not. Can the app give me another one?"* Therefore, SalienTrack should provide dynamically selected concise feedback, with the option to expand to show more items to accommodate mistakes in saliency selection.

**Preference for SalienTrack Auto and Semi-Auto modes over Manual.** Participants disliked Manual feedback the most due to its tediousness and their lack of knowledge. P6 explained that *"if I have to manually log information, I wouldn't use it for long time. I want automated information."* P9 said *"if I eat in a restaurant, I've no idea how much, like sugar, salt, oil they put into the food."* However, after some prompting, participants could see some value with Manual feedback. P2 *"like[d] the idea of Manual version that it forces me to think about the food. But I am not nutrition expert, so I don't know if I can estimate the value correctly if I manually log."* Therefore, SalienTrack should provide Semi-Auto feedback, with item information shown automatically most of the time, and sometimes manually elicited.

**Desire for feedback of all meals.** With SalienTrack, some meals were omitted from feedback due to low predicted informativeness. However, participants generally were eager to see feedback for all meals and were curious to know how some meals were predicted to be low informativeness. P3 wanted to reap the fruits of his efforts and remarked that *"I've taken the photos, so it makes no sense to provide no information to me at all. I can ignore the information if I don't have time, but it's good to have the information."* P2 understood the benefit of not reviewing all meals, remarking *"it's OK that the App says no feedback for this meal if there is really no interesting information."* P1's curiosity was piqued: *"When the App says there is no interesting information of this meal for me, this makes me get interested in why the App thinks it's not interesting?"* Therefore, SalienTrack should have the option to show low-informative meals on demand. We believe that interest in viewing all meals will wane over time, so a longitudinal study is needed to evaluate the usefulness of limiting salient moments (when).

**Need for contextual, valence-based and actionable insights.** In addition to viewing factual information, participants wanted the feedback to be contextualized to their health goals [11] and include action plans [2]. This agrees with our earlier findings on interpreting feedback by the valence of the meal (Section 4.5). P7 wanted to *"categorize the information by positive and negative. Then I know what I am doing good, what I can improve."* P4 felt that *"some information is vague for me. I don't know what I am supposed to do given the information. I like clear suggestions."* Therefore, SalienTrack could be combined with a healthiness prediction model to indicate meals and features that support or undermine healthiness, and combined with a recommender system for action plans. For example, when stating "this meal was deep fried", SalienTrack could contextualize that this was an "unhealthy cooking method with much fat", which is harmful towards a low-fat diet, and suggest to "consider baking instead".

## 7 DISCUSSION

We have answered our research questions: *RQ1) What information is salient in feedback? RQ2) How to provide salient information in feedback?* We summarize the evidence for salient feedback, discuss its implications for informativeness in self-tracking, how adding model explainability expands opportunities for feedback experience, and how to generalize our saliency approach to other self-tracking activities.

### 7.1 Dimensions for Salient Feedback in Self-Tracking

We examined the need for, provision of, and usefulness of salient feedback in data collection, modeling, and formative studies. Table 6 summarizes our findings along the four dimensions of the SalienTrack framework. Therefore, to sustain engagement, salient feedback should be provided occasionally, with concise details, with rationales supportive of or antagonistic to the user's health goals, and with diverse feedback modes.

Table 6: Evidence to support the usefulness towards saliency for each dimension in the SalienTrack framework.

| Dimension | Evidence for Support |
|---|---|
| When | *Data collection study:* Informativeness ratings per meal were varied across high and low ratings.<br>*Modeling study:* Gradient Boosted Tree can accurately predict informativeness (F1 Score = 0.74, 0.84).<br>*Formative study:* Participants wanted feedback for all meals initially, but agreed seeing fewer is less tedious. |
| Which | *Data collection study:* Found that participants reflected on diet behaviors more than nutrition knowledge, and mentioned macronutrients more and ingredients less in negative reflections than in positive ones.<br>*Modeling study:* Some features are more salient than others for each prediction instance, and salient features dynamically change for each feedback instance.<br>*Formative study:* Found that participants prefer concise feedback, appreciated dynamic salient selection, and preferred historical information over detailed nutrition knowledge. |
| Why | *Data collection study:* Found that participants explained their diet behaviors using contextual information.<br>*Modeling study:* Anchor rules learned decision boundaries to reveal counterfactual changes to lead to different prediction outcomes of informativeness.<br>*Formative study:* Found that participants wanted to relate feedback items to their healthiness goals. |
| How | *Data collection study:* Found that participants rated Auto feedback as more informative than from Manual.<br>*Modeling study:* Auto and Manual prediction models may select different salient features for the same instance, suggesting that informativeness depends on feedback mode.<br>*Formative study:* Found that participants prefer Auto feedback than Manual due to the latter's tediousness, but appreciated occasional Manual feedback for deeper reflection (i.e., most preferred SalienTrack-SemiAuto). |

### 7.2 Limitations and Scope of Data Collection for Salient Feedback

We discuss limitations in our data collection study. 1) We had limited the feedback to nutrition information to limit user burden, and excluded other contextual information, such as events, places, and people [52]. The excluded features may occasionally be more salient than nutrition knowledge or diet behaviors, which future work should investigate. 2) We had also limited feedback sessions to weekly intervals, but future work can explore how much saliency is beneficial for varying feedback frequencies at every meal [47] or once per day [76], and across years [84], which will have differing total logged data amounts and different reflection patterns [37]. 3) There was a high drop-off rate in the data collection study, though this is a common issue in health behavior change studies [26]. Our persistent users may be biased towards engagement and rate informativeness higher than average.

### 7.3 Machine Learning Predictions and Explanations for Self-Tracking Feedback

Many machine learning techniques have been proposed to address burdens in collection and reflection for self-tracking. For data collection, models can automatically detect and infer events, such as physical activity [76], sleep [49], stress [1], and eating [9,87,91]. For feedback reflection, models can predict opportunities, such as recommending context-aware actions [76] and providing just-in-time adaptive interventions (JITAI) [53,70], finding associations between macronutrients blood glucose [65], and identifying disengaged users [55]. We add to the latter body of work by providing salient information subsets based on the potential to learn from feedback.

As more sophisticated models are used to support self-tracking and feedback, there is more need for explainable AI to support human reasoning goals [89]. For self-tracking, instead of generating explanations for data scientists to debug the models, it is important to explain to domain experts, such as diabetes educators [65] and public health officials [55], and explain to consumers to persuade healthier behaviors [33]. We have leveraged Anchor rules to justify to the self-tracker why they should pay attention to the feedback. Our formative study informed how combining with health prediction modeling can provide more persuasive and explicit goal-oriented feedback.

Currently, the saliency selection of SalienTrack is purely data-driven based on end-user data, but the designer or dietitian may choose to prioritize some features. For example, we found that showing feedback on ingredients tends to produce positive valence reflections. Future work can encode this requirement as a prediction *prior* to using machine learning regularization to prioritize features for positive valence. To increase the clinical relevance of the salient selected features, dietitians can also specify higher priority factors (e.g., mentioning the vegetables food group and fiber macronutrient), and these can be added as regularization terms to penalize their omission.

**7.4  Generalizing SalienTrack to other Self-Tracking**

We have demonstrated SalienTrack for selectively providing concise and dynamic feedback for meal self-tracking. Applying the SalienTrack technique involves multiples steps in two phases. We summarize them to help replicate the approach. First, conduct a usage trial of a self-tracking app with feedback intervention, similar to diet tracking trials (e.g., [11,30,48]) or studies on personal informatics (e.g., [30,35,54]). However, other than just recording activity or providing feedback, also survey the users on their perceived informativeness of the feedback for each tracked event, and for each information item. The data collection step is necessary to identify salient information due to differing application (e.g., healthy lifestyles or chronic disease management), cultural, and population contexts. For example, including carbohydrates is important for ethnic Indians with higher diabetes risk [66], and including sodium intake can help Japanese users manage their stroke risk [69]). Finally, extract features from quantitatively and qualitatively analyzing the perceived informativeness ratings and rationales We expect that historical temporal features that we extracted in our food logging use case to also be relevant for other applications.

The second phase involves engineering the AI system and app interface for salient feedback: 1) Train a machine learning model to predict informativeness. 2) Implement explanations using SHAP for saliency and Anchors for rule reasons. 3) Implement the salient feedback by displaying or eliciting only the top-ranked features based on the model prediction confidences of the Auto and Manual models, respectively. We only demonstrated the feedback UI with simple lists and form widgets like in [30], but other pictorials or visualizations can be considered (e.g., [4,36]). Note that saliency and informativeness predictions depend on the UI format; with better interactive visualizations, more informativeness may be higher, and tediousness may be smaller, so modeling results may be different.

The SalienTrack technique can also be applied to other behaviors, such as physical activity, sleep, and savings. We discuss applying SalienTrack to sleep tracking [49] with multiple steps. Step 0) track contextual information, such as sleep time, coffee intake, time between exercise and sleep, room temperature, feedback informativeness rating, 1) train a model to predict *when* a user will learn from sleep episode feedback (whether about good or poor sleep), 2) apply SHAP to select *which* features to include in the feedback, and 3) determine *how* to present the feedback, e.g., manually ask about temperature comfort, or automatically show amount of movement during sleep). To contextualize feedback with goals, such as sleep quality, a prediction model should be trained with the contextual features and Anchor rule explanations provided to explain *why* the context helped or harmed sleep.

Some users may be frustrated with needing to do manual recording work despite the app having the automatic prediction capability. Nevertheless, the proposed ability to switch between Manual and Auto feedback in SalienTrack can mitigate users "losing the habit" and consequently abandoning self-tracking [31] by substituting forgotten logs with manual reflections. Furthermore, there is a technical benefit for occasionally switching the feedback from Auto to Manual. The manual feedback can be used for active learning for the machine, i.e., to prompt the user to provide annotations when the machine learning model's prediction confidence is low. This provides labeling data to help the model improve its accuracy and improve the feasibility towards Auto feedback. Hence, SalienTrack can be used for active learning [71] for the user and the machine [27].

In general, SalienTrack can be used if automatic feedback can be obtained for a tracked behavior [20]. Data collection need not be automated, as is the case for mobile photo-based food logging. Applications where automatic inference remains elusive include capturing the social context or background of an activity (e.g., overeating due to attending a party), and feeling bloated after eating triggering foods [24]. For these applications where manual elicitation is needed for assessment, users will already be confounded with being burdened to reflect. In such cases, applying decision-theoretic models can help to mitigate repeated elicitations by accounting for their costs [53,78].

## 8 CONCLUSION

We have studied what users find salient in self-tracking feedback and found differences in perceived informativeness across logged meals, and for different nutrition, assessment, diet behavioral, and contextual information. Applying these insights, we quantified the informativeness in self-tracking and proposed the SalienTrack framework that defines the saliency of *when*, with *which* information, *why*, and *how* to provide feedback. We implemented a machine learning model with explanations to predict the informativeness of feedback at each meal event, and explain the most salient information for users to learn. Our formative study showed the usefulness of SalienTrack to provide concise, dynamic feedback. SalienTrack demonstrates semi-automatic feedback based on informativeness, and expands opportunities to make feedback more concise and engaging.

**REFERENCES**

1. Phil Adams, Mashfiqui Rabbi, Tauhidur Rahman, Mark Matthews, Amy Voida, Geri Gay, Tanzeem Choudhury, and Stephen Voida. 2014. Towards personal stress informatics: Comparing minimally invasive techniques for measuring daily stress in the wild. *Proceedings - PERVASIVEHEALTH 2014: 8th International Conference on Pervasive Computing Technologies for Healthcare*: 72–79. https://doi.org/10.4108/icst.pervasivehealth.2014.254959
2. Elena Agapie, Bonnie Chinh, Laura R Pina, Diana Oviedo, Molly C Welsh, Gary Hsieh, and Sean Munson. 2018. Crowdsourcing Exercise Plans Aligned with Expert Guidelines and Everyday Constraints. In *CHI 2018*, 324.
3. Beth Archer-Kuhn, Natalie R. Beltrano, Judith Hughes, Michael Saini, and Dora Tam. 2021. Recruitment in response to a pandemic: pivoting a community-based recruitment strategy to facebook for hard-to-reach populations during COVID-19. *International Journal of Social Research Methodology* 00, 00: 1–12. https://doi.org/10.1080/13645579.2021.1941647
4. Amid Ayobi, Paul Marshall, and Anna L. Cox. 2020. Trackly: A Customisable and Pictorial Self-Tracking App to Support Agency in Multiple Sclerosis Self-Care. In *CHI 2020*, 1–15. https://doi.org/10.1145/3313831.3376809
5. Amid Ayobi, Tobias Sonne, Paul Marshall, and Anna L. Cox. 2018. Flexible and Mindful Self-Tracking: Design Implications from Paper Bullet Journals. In *CHI 2018*, 1–14. https://doi.org/10.1145/3173574.3173602
6. Mary L. Barreto, Agnieszka Szóstek, Evangelos Karapanos, Nuno J. Nunes, Lucas Pereira, and Filipe Quintal.

2014. Understanding families' motivations for sustainable behaviors. *Computers in Human Behavior* 40, November: 6–15. https://doi.org/10.1016/j.chb.2014.07.042
7. Kenneth Bates, Scot Burton, Kyle Huggins, and Elizabeth Howlett. 2011. Battling the bulge: menu board calorie legislation and its potential impact on meal repurchase intentions. *Journal of Consumer Marketing* 28, 2: 104–113. https://doi.org/10.1108/07363761111115944
8. Eric P.S. Baumer. 2015. Reflective Informatics: Conceptual Dimensions for Designing Technologies of Reflection. In *CHI 2015*, 585–594. https://doi.org/10.1145/2702123.2702234
9. Abdelkareem Bedri, Richard Li, Malcolm Haynes, Raj Prateek Kosaraju, Ishaan Grover, Temiloluwa Prioleau, Min Yan Beh, Mayank Goel, Thad Starner, and Gregory Abowd. 2017. EarBit: Using Wearable Sensors to Detect Eating Episodes in Unconstrained Environments. *Proc. ACM Interact. Mob. Wearable Ubiquitous Technol.* 1, 3: 37:1--37:20. https://doi.org/10.1145/3130902
10. H. Beyer and K Holtzblatt. 1998. *Contextual design: defining customer-centered systems*. Morgan Kaufmann.
11. Johnna Blair, Yuhan Luo, Ning F. Ma, Sooyeon Lee, and Eun Kyoung Choe. 2018. OneNote Meal: A Photo-Based Diary Study for Reflective Meal Tracking. In *AMIA Annual Symposium Proceedings*, 252–261.
12. A. J.Bernheim Brush, Evgeni Filippov, Danny Huang, Jaeyeon Jung, Ratul Mahajan, Frank Martinez, Khurshed Mazhar, Amar Phanishayee, Arjmand Samuel, James Scott, and Rayman Preet Singh. 2013. Lab of things: A platform for conducting studies with connected devices in multiple homes. *UbiComp 2013 Adjunct - Adjunct Publication of the 2013 ACM Conference on Ubiquitous Computing*: 35–38. https://doi.org/10.1145/2494091.2494100
13. A.J. Bernheim Brush, Jaeyeon Jung, Ratul Mahajan, and James Scott. 2012. HomeLab: Shared Infrastructure for Home Technology Field Studies. In *UbiComp'12*, 1108–1113. https://doi.org/10.1145/2486001.2491701
14. Marissa Burgermaster, Krzysztof Z. Gajos, Patricia Davidson, and Lena Mamykina. 2017. The role of explanations in casual observational learning about nutrition. *CHI 2017* 2017-May: 4097–4108. https://doi.org/10.1145/3025453.3025874
15. S. Butscher, Y. Wang, J. Mueller, K. Ziesemer, K. Villinger, D. Wahl, L. Koenig, G. Sproesser, B. Renner, H.T. Schupp, and H. Reiterer. 2016. Lightweight visual data analysis on mobile devices - Providing self-monitoring feedback. In *CEUR Workshop Proceedings*.
16. Weiwen Chai and Michael Liebman. 2005. Effect of different cooking methods on vegetable oxalate content. *Journal of Agricultural and Food Chemistry* 53, 8: 3027–3030. https://doi.org/10.1021/jf048128d
17. Jing Jing Chen, Chong Wah Ngo, Fu Li Feng, and Tat Seng Chua. 2018. Deep understanding of cooking procedure for cross-modal recipe retrieval. *MM 2018 - Proceedings of the 2018 ACM Multimedia Conference*: 1020–1028. https://doi.org/10.1145/3240508.3240627
18. Jingjing Chen and Chong-wah Ngo. 2016. Deep-based Ingredient Recognition for Cooking Recipe Retrieval. In *Proceedings of the 2016 ACM on Multimedia Conference - MM '16*, 32–41. https://doi.org/10.1145/2964284.2964315
19. Tianqi Chen and Carlos Guestrin. 2016. XGBoost: A Scalable Tree Boosting System. In *KDD '16*, 785–794. https://doi.org/10.1145/2939672.2939785
20. Eun Kyoung Choe, Saeed Abdullah, Mashfiqui Rabbi, Edison Thomaz, Daniel A Epstein, Matthew Kay, Gregory D Abowd, James Fogarty, Bongshin Lee, Mark Matthews, and Julie A Kientz. 2017. Semi-Automated Tracking: A Balanced Approach for Self-Monitoring Applications. *IEEE Pervasive Computing* 16, 1: 74–84. https://doi.org/10.1109/MPRV.2017.18
21. Eun Kyoung Choe, Bongshin Lee, and M. C. Schraefel. 2015. Characterizing Visualization Insights from Quantified Selfers' Personal Data Presentations. *IEEE Computer Graphics and Applications* 35, 4: 28–37. https://doi.org/10.1109/MCG.2015.51
22. Eun Kyoung Choe, Bongshin Lee, Haining Zhu, and Nathalie Henry Riche. 2017. Understanding self-reflection: How people reflect on personal data through visual data exploration. *Pervasive Health '17*: 173–182.

https://doi.org/10.1145/3154862.3154881
23. Chia-Fang Chung, Elena Agapie, Jessica Schroeder, Sonali Mishra, James Fogarty, and Sean A. Munson. 2017. When Personal Tracking Becomes Social: Examining the Use of Instagram for Healthy Eating. In *CHI 2017*, 1674–1687. https://doi.org/10.1145/3025453.3025747
24. Chia-Fang Chung, Qiaosi Wang, Jessica Schroeder, Allison Cole, Jasmine Zia, James Fogarty, and Sean A. Munson. 2019. Identifying and Planning for Individualized Change: Patient-Provider Collaboration Using Lightweight Food Diaries in Healthy Eating and Irritable Bowel Syndrome. *Proceedings of the ACM on Interactive, Mobile, Wearable and Ubiquitous Technologies* 3, 1: 1–27. https://doi.org/10.1145/3314394
25. Chia Fang Chung, Kristin Dew, Allison Cole, Jasmine Zia, James Fogarty, Julie A. Kientz, and Sean A. Munson. 2016. Boundary negotiating artifacts in personal informatics: Patient-provider collaboration with patient-generated data. *Proceedings of the ACM Conference on Computer Supported Cooperative Work, CSCW* 27: 770–786. https://doi.org/10.1145/2818048.2819926
26. James Clawson, Jessica A. Pater, Andrew D. Miller, Elizabeth D. Mynatt, and Lena Mamykina. 2015. No longer wearing: Investigating the abandonment of personal health-Tracking technologies on craigslist. *UbiComp 2015 - Proceedings of the 2015 ACM International Joint Conference on Pervasive and Ubiquitous Computing*: 647–658. https://doi.org/10.1145/2750858.2807554
27. David A. Cohn, Zoubin Ghahramani, and Michael I. Jordan. 1996. Active Learning with Statistical Models. *Journal of Arti?cial Intelligence Research* 4: 129–145.
28. Sunny Consolvo, Predrag Klasnja, David W McDonald, and James A Landay. 2009. Goal-setting Considerations for Persuasive Technologies That Encourage Physical Activity. In *Proceedings of the 4th International Conference on Persuasive Technology* (Persuasive '09), 8:1--8:8. https://doi.org/10.1145/1541948.1541960
29. Sunny Consolvo, David W McDonald, Tammy Toscos, Mike Y Chen, Jon Froehlich, Beverly Harrison, Predrag Klasnja, Anthony LaMarca, Louis LeGrand, Ryan Libby, Ian Smith, and James A Landay. 2008. Activity Sensing in the Wild: A Field Trial of Ubifit Garden. In *Proceedings of the SIGCHI Conference on Human Factors in Computing Systems* (CHI '08), 1797–1806. https://doi.org/10.1145/1357054.1357335
30. Felicia Cordeiro, Elizabeth Bales, Erin Cherry, and James Fogarty. 2015. Rethinking the mobile food journal: Exploring opportunities for lightweight photo-based capture. In *CHI 2015*, 3207–3216. https://doi.org/10.1145/2702123.2702154
31. Felicia Cordeiro, Daniel A. Epstein, Edison Thomaz, Elizabeth Bales, Arvind K. Jagannathan, Gregory D. Abowd, and James Fogarty. 2015. Barriers and negative nudges: Exploring challenges in food journaling. In *Conference on Human Factors in Computing Systems - Proceedings*, 1159–1162. https://doi.org/10.1145/2702123.2702155
32. LOUIS P. CUSELLA. 1982. the Effects of Source Expertise and Feedback Valence on Intrinsic Motivation. *Human Communication Research* 9, 1: 17–32. https://doi.org/10.1111/j.1468-2958.1982.tb00680.x
33. Mauro Dragoni, Ivan Donadello, and Claudio Eccher. 2020. Explainable AI Meets Persuasiveness: Translating Reasoning Results Into Behavioral Change Advice. *Artificial Intelligence in Medicine*: 101840. https://doi.org/10.1016/j.artmed.2020.101840
34. Takumi Ege and Keiji Yanai. 2017. Image-based food calorie estimation using knowledge on food categories, ingredients and cooking directions. *Thematic Workshops 2017 - Proceedings of the Thematic Workshops of ACM Multimedia 2017, co-located with MM 2017*: 367–375. https://doi.org/10.1145/3126686.3126742
35. Daniel A. Epstein, Felicia Cordeiro, James Fogarty, Gary Hsieh, and Sean A. Munson. 2016. Crumbs: Lightweight daily food challenges to promote engagement and mindfulness. In *Conference on Human Factors in Computing Systems - Proceedings*, 5632–5644. https://doi.org/10.1145/2858036.2858044
36. Daniel A Epstein, Felicia Cordeiro, Elizabeth Bales, James Fogarty, and Sean A Munson. 2014. Taming Data Complexity in Lifelogs: Exploring Visual Cuts of Personal Informatics Data. In *DIS 2014*. https://doi.org/10.1145/2598510.2598558/

37. Daniel A Epstein, Parisa Eslambolchilar, Judy Kay, Jochen Meyer, and Sean A Munson. 2021. Opportunities and challenges for long-term tracking. In *Advances in Longitudinal HCI Research*. Springer, 177–206.
38. Gunther Eysenbach. 2005. The law of attrition. *Journal of Medical Internet Research* 7, 1: 1–9. https://doi.org/10.2196/jmir.7.1.e11
39. Shaobo Fang, Chang Liu, Fengqing Zhu, Edward J. Delp, and Carol J. Boushey. 2015. Single-View Food Portion Estimation Based on Geometric Models. In *2015 IEEE International Symposium on Multimedia*. https://doi.org/10.1109/ISM.2015.67
40. Jon Froehlich, Tawanna Dillahunt, Predrag Klasnja, Jennifer Mankoff, Sunny Consolvo, Beverly Harrison, and James A. Landay. 2009. UbiGreen: Investigating a Mobile Tool for Tracking and Supporting Green Transportation Habits. In *CHI'09*, 1043–1052.
41. Federica Gerina, Silvia M. Massa, Francesca Moi, Diego Reforgiato Recupero, and Daniele Riboni. 2020. Recognition of cooking activities through air quality sensor data for supporting food journaling. *Human-centric Computing and Information Sciences* 10, 1. https://doi.org/10.1186/s13673-020-00235-9
42. Barney G. Glaser and Anselm L. Strauss. 2006. *The Discovery of grounded theory: strategies for qualitative research*. AldineTransaction.
43. Bruce Hanington and Bella Martin. 2021. *Universal methods of design: 100 ways to research complex problems, develop innovative ideas, and design effective solutions*. Rockport Publishers.
44. Jiangpeng He, Zeman Shao, Janine Wright, Deborah Kerr, Carol Boushey, and Fengqing Zhu. 2020. Multi-Task Image-Based Dietary Assessment for Food Recognition and Portion Size Estimation. *2020 IEEE Conference on Multimedia Information Processing and Retrieval (MIPR)*: 49–54. https://doi.org/10.1109/MIPR49039.2020.00018
45. David R Jacobs and Linda C Tapsell. 2007. Food, Not Nutrients, Is the Fundamental Unit in Nutrition. *Nutrition Reviews* 65, 10: 439–450. https://doi.org/10.1111/j.1753-4887.2007.tb00269.x
46. M Juárez, S Failla, A Ficco, F Peña, C Avilés, and O Polvillo. 2010. Buffalo meat composition as affected by different cooking methods. *Food and Bioproducts Processing* 88, 2: 145–148. https://doi.org/10.1016/j.fbp.2009.05.001
47. Jisu Jung, Lyndal Wellard-Cole, Colin Cai, Irena Koprinska, Kalina Yacef, Margaret Allman-Farinelli, and Judy Kay. 2020. Foundations for Systematic Evaluation and Benchmarking of a Mobile Food Logger in a Large-scale Nutrition Study. *Proceedings of the ACM on Interactive, Mobile, Wearable and Ubiquitous Technologies* 4, 2. https://doi.org/10.1145/3397327
48. Yi-Chin Kato-Lin, Vibhanshu Abhishek, Julie S. Downs, and Rema Padman. 2016. Food for Thought: The Impact of m-Health Enabled Interventions on Eating Behavior. *SSRN Electronic Journal*. https://doi.org/10.2139/ssrn.2736792
49. Matthew Kay, Eun Kyoung Choe, Jesse Shepherd, Benjamin Greenstein, Nathaniel Watson, Sunny Consolvo, and Julie A Kientz. 2012. Lullaby: A Capture & Access System for Understanding the Sleep Environment. In *Proceedings of the 2012 ACM Conference on Ubiquitous Computing* (UbiComp '12), 226–234. https://doi.org/10.1145/2370216.2370253
50. Rafal Kocielnik, Lillian Xiao, Daniel Avrahami, and Gary Hsieh. 2018. Reflection Companion: A Conversational System for Engaging Users in Reflection on Physical Activity. *Proceedings of the ACM on Interactive, Mobile, Wearable and Ubiquitous Technologies* 2, 2: 1–26. https://doi.org/10.1145/3214273
51. Ian Li, Anind Dey, and Jodi Forlizzi. 2010. A Stage-Based Model of Personal Informatics Systems. In *CHI 2010*, 557–566.
52. Ian Li, Anind K Dey, and Jodi Forlizzi. 2012. Using Context to Reveal Factors That Affect Physical Activity. *ACM Trans. Comput.-Hum. Interact.* 19, 1: 7:1--7:21. https://doi.org/10.1145/2147783.2147790
53. Peng Liao, Kristjan Greenewald, Predrag Klasnja, and Susan Murphy. 2020. Personalized HeartSteps: A reinforcement learning algorithm for optimizing physical activity. *Proc. ACM Interact. Mob. Wearable*

*Ubiquitous Technol.* 4.

54. Brian Y. Lim, Xinni Chng, and Shengdong Zhao. 2017. Trade-off between automation and accuracy in mobile photo recognition food logging. In *Chinese CHI 2017*, 53–59. https://doi.org/10.1145/3080631.3080640
55. Brian Y. Lim, Judy Kay, and Weilong Liu. 2019. How Does a Nation Walk? Interpreting Large-Scale Step Count Activity with Weekly Streak Patterns. *Proceedings of the ACM on Interactive, Mobile, Wearable and Ubiquitous Technologies* 3, 2: 1–46. https://doi.org/10.1145/3328928
56. Brian Y Lim and Anind K Dey. 2009. Assessing Demand for Intelligibility in Context-Aware Applications. In *UbiComp 2009*, 195–204.
57. Penelope Lockwood, Pamela Sadler, Keren Fyman, and Sarah Tuck. 2004. To do or not to do: Using positive and negative role models to harness motivation. *Social Cognition* 22, 4: 422–450. https://doi.org/10.1521/soco.22.4.422.38297
58. Kai Lukoff, Yuan Zhuang, Brian Y Lim, and Taoxi Li. 2018. TableChat: Mobile Food Journaling to Facilitate Family Support for Healthy Eating. *Proceedings of the ACM on Human-Computer Interaction* 2, CSCW. https://doi.org/10.1145/3274383
59. Scott M Lundberg and Su-In Lee. 2017. A Unified Approach to Interpreting Model Predictions. In *31st Conference on Neural Information Processing Systems (NIPS 2017)*. Retrieved May 27, 2019 from https://github.com/slundberg/shap
60. Yuhan Luo, Young-ho Kim, and Bongshin Lee. 2021. FoodScrap: Promoting Rich Data Capture and Reflective Food Journaling Through Speech Input. In *DIS'21*, 606–618.
61. Yuhan Luo, Peiyi Liu, and Eun Kyoung Choe. 2019. Co-Designing Food Trackers with Dietitians: Identifying Design Opportunities for Food Tracker Customization. In *CHI 2019*, 1–13. https://doi.org/10.1145/3290605.3300822
62. L. Kathleen Mahan and Sylvia Escott-Stump. 2016. *Krause's food & the nutrition care process-e-book*. Elsevier Health Sciences.
63. Lakmal Meegahapola, Salvador Ruiz-Correa, Viridiana Del Carmen Robledo-Valero, Emilio Ernesto Hernandez-Huerfano, Leonardo Alvarez-Rivera, Ronald Chenu-Abente, and Daniel Gatica-Perez. 2021. One More Bite? Inferring Food Consumption Level of College Students Using Smartphone Sensing and Self-Reports. *Proceedings of the ACM on Interactive, Mobile, Wearable and Ubiquitous Technologies* 5, 1: 1–28. https://doi.org/10.1145/3448120
64. Mark Mirtchouk, Christopher Merck, and Samantha Kleinberg. 2016. Automated estimation of food type and amount consumed from body-worn audio and motion sensors. In *UbiComp 2016 - Proceedings of the 2016 ACM International Joint Conference on Pervasive and Ubiquitous Computing*, 451–462. https://doi.org/10.1145/2971648.2971677
65. Elliot G Mitchell, Elizabeth M Heitkemper, Marissa Burgermaster, Matthew E Levine, Maria L Hwang, Andrea Cassells, Jonathan N Tobin, Esteban G Tabak, David J Albers, Arlene M Smaldone, and Lena Mamykina. 2021. From Reflection to Action : Combining Machine Learning with Expert Knowledge for Nutrition Goal Recommendations. In *CHI'21*.
66. V Mohan. 2004. Why are Indians more prone to diabetes? *The Journal of the Association of Physicians of India* 52: 468–474.
67. James W. Moore. 2016. What is the sense of agency and why does it matter? *Frontiers in Psychology* 7, AUG: 1–9. https://doi.org/10.3389/fpsyg.2016.01272
68. Austin Myers, Nick Johnston, Vivek Rathod, Anoop Korattikara, Alex Gorban, Nathan Silberman, Sergio Guadarrama, George Papandreou, Jonathan Huang, and Kevin Murphy. 2015. Im2Calories: Towards an automated mobile vision food diary. *Proceedings of the IEEE International Conference on Computer Vision* 2015 Inter, December: 1233–1241. https://doi.org/10.1109/ICCV.2015.146
69. Chisato Nagata, Naoyoshi Takatsuka, Natsuki Shimizu, and Hiroyuki Shimizu. 2004. Sodium intake and risk of

death from stroke in Japanese men and women. *Stroke* 35, 7: 1543–1547.
70. Inbal Nahum-Shani, Shawna N. Smith, Bonnie J. Spring, Linda M. Collins, Katie Witkiewitz, Ambuj Tewari, and Susan A. Murphy. 2016. Just-in-Time Adaptive Interventions (JITAIs) in Mobile Health: Key Components and Design Principles for Ongoing Health Behavior Support. *Annals of Behavioral Medicine*: 1–17. https://doi.org/10.1007/s12160-016-9830-8
71. Hannele Niemi. 2002. Active learning - A cultural change needed in teacher education and schools. *Teaching and Teacher Education* 18, 7: 763–780. https://doi.org/10.1016/S0742-051X(02)00042-2
72. Jon Noronha, Eric Hysen, Haoqi Zhang, and Krzysztof Z. Gajos. 2011. PlateMate: Crowdsourcing nutrition analysis from food photographs. *UIST'11 - Proceedings of the 24th Annual ACM Symposium on User Interface Software and Technology*: 1–11. https://doi.org/10.1145/2047196.2047198
73. Homin Park, Homanga Bharadhwaj, and Brian Y. Lim. 2019. Hierarchical Multi-Task Learning for Healthy Drink Classification. *Proceedings of the International Joint Conference on Neural Networks* 2019-July. https://doi.org/10.1109/IJCNN.2019.8851796
74. Homin Park, Abhinav Ramesh Kashyap, Zhenkai Wang, and Brian Y Lim. 2018. Biases in Food Photo Taking Behavior. In *Designing Recipes for Digital Food Futures, a CHI workshop*. Retrieved November 29, 2019 from http://datamaterialities.org/chi2018workshop.html
75. Rebecca Passonneau. 2006. Measuring agreement on set-valued items (MASI) for semantic and pragmatic annotation. *Proceedings of the 5th International Conference on Language Resources and Evaluation, LREC 2006*: 831–836.
76. Mashfiqui Rabbi, Min Hane Aung, Mi Zhang, and Tanzeem Choudhury. 2015. MyBehavior: automatic personalized health feedback from user behaviors and preferences using smartphones. In *Proceedings of the 2015 ACM International Joint Conference on Pervasive and Ubiquitous Computing (UbiComp '15 )*, 707–718. https://doi.org/10.1145/2750858.2805840
77. Marco Tulio Ribeiro and Carlos Guestrin. 2018. Anchors : High-Precision Model-Agnostic Explanations. In *The Thirty-Second AAAI Conference on Artificial Intelligence (AAAI-18)*, 1527–1535.
78. Stephanie Rosenthal, Anind K. Dey, and Manuela Veloso. 2011. Using decision-theoretic experience sampling to build personalized mobile phone interruption models. In *International Conference on Pervasive Computing*, 170–187. https://doi.org/10.1007/978-3-642-21726-5_11
79. Doyen Sahoo, Wang Hao, Shu Ke, Wu Xiongwei, Hung Le, Palakorn Achananuparp, Ee Peng Lim, and Steven C.H. Hoi. 2019. FoodAI: Food image recognition via deep learning for smart food logging. In *Proceedings of the ACM SIGKDD International Conference on Knowledge Discovery and Data Mining*, 2260–2268. https://doi.org/10.1145/3292500.3330734
80. Jessica Schroeder, Jane Hoffswell, Chia Fang Chung, James Fogarty, Sean Munson, and Jasmine Zia. 2017. Supporting patient-provider collaboration to identify individual triggers using food and symptom journals. *Proceedings of the ACM Conference on Computer Supported Cooperative Work, CSCW*: 1726–1739. https://doi.org/10.1145/2998181.2998276
81. Julia Schwarz, Jennifer Mankoff, and H. Scott Matthews. 2009. Reflections of everyday activities in spending data. *Conference on Human Factors in Computing Systems - Proceedings*: 1737–1740. https://doi.org/10.1145/1518701.1518968
82. Ralf Schwarzer. 2008. Modeling health behavior change: How to predict and modify the adoption and maintenance of health behaviors. *Applied Psychology* 57, 1: 1–29. https://doi.org/10.1111/j.1464-0597.2007.00325.x
83. Lucas M Silva and Daniel A Epstein. 2021. Investigating Preferred Food Description Practices in Digital Food Journaling. In *DIS'21*, 589–605.
84. Lie Ming Tang and Judy Kay. 2017. Harnessing long term physical activity data - how long-term trackers use data and how an adherence-based interface supports new insights. *Preceedings of the ACM on Interactive,*

*Mobile, Wearable and Ubiquitous Technologies* 1, 26. https://doi.org/10.1145/3090091
85. Tauhidur Rahman, Mary Czerwinski, Ran Gilad-Bachrach, and Paul Johns. 2016. Predicting "About-to-Eat" Moments for Just-in-Time Eating Intervention Tauhidur. In *Proceedings of the 6th International Conference on Digital Health Conference*, 141–150. https://doi.org/10.1145/2896338.2896359
86. The Culinary Institute of America. 2016. 3 Types of Cooking Methods and the Foods That Love Them. Retrieved from https://blog.ciachef.edu/3-types-cooking-methods-foods-love/
87. Edison Thomaz, Irfan Essa, and Gregory D Abowd. 2015. A Practical Approach for Recognizing Eating Moments with Wrist-mounted Inertial Sensing. In *Proceedings of the 2015 ACM International Joint Conference on Pervasive and Ubiquitous Computing* (UbiComp '15), 1029–1040. https://doi.org/10.1145/2750858.2807545
88. Elizabeth Salerno Valdez and Aline Gubrium. 2020. Shifting to Virtual CBPR Protocols in the Time of Corona Virus/COVID-19. *International Journal of Qualitative Methods* 19: 1–9. https://doi.org/10.1177/1609406920977315
89. Danding Wang, Qian Yang, Ashraf Abdul, and Brian Y Lim. 2019. Designing Theory-Driven User-Centric Explainable AI. In *CHi '19*. https://doi.org/10.1145/3290605.3300831
90. Yunlong Wang, Le Duan, Jens Mueller, Simon Butscher, and Harald Reiterer. 2016. "Fingerprints": Detecting meaningful moments for mobile health intervention. In *Proceedings of the 18th International Conference on Human-Computer Interaction with Mobile Devices and Services Adjunct, MobileHCI 2016*, 1085–1088. https://doi.org/10.1145/2957265.2965006
91. Shibo Zhang, Yuqi Zhao, Dzung Tri Nguyen, Runsheng Xu, Sougata Sen, Josiah Hester, and Nabil Alshurafa. 2020. NeckSense: A Multi-Sensor Necklace for Detecting Eating Activities in Free-Living Conditions. *Proceedings of the ACM on Interactive, Mobile, Wearable and Ubiquitous Technologies* 4, 2: 1–26. https://doi.org/10.1145/3397313

**APPENDICES**

**A SURVEYS**

This appendix shows the different surveys used in the self-tracking informativeness study.

**A.1 Screening Survey**

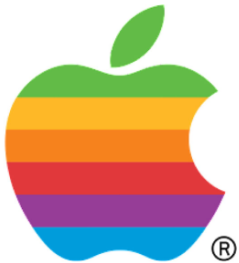

3. Describe in a sentence what you see in the photo *

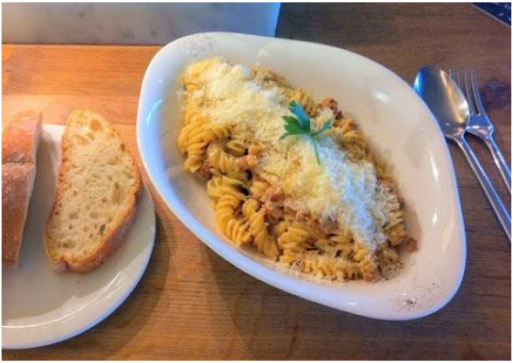

_______________________________________________
_______________________________________________
_______________________________________________
_______________________________________________
_______________________________________________

4. Enter your mturk worker ID *

    Please provide your mturk worker ID here. This will also be used for the survey code

    _______________________________________________

5. What is your gender? *

    *Mark only one oval.*

    ◯ Male
    ◯ Female
    ◯ Other: _______________________________

This content is neither created nor endorsed by Google.

Google Forms

### A.2 Pre-Survey

## Pre-survey

Thank you for participating in our study! Please carefully read questions and descriptions before answering them. Please feel free to contact [REDACTED] if you encounter any issue.

* Required

Please tell us a little about yourself

1. How old are you? *

    *Mark only one oval.*

    ◯ 18-24 years old
    ◯ 25-34 years old
    ◯ 35-44 years old
    ◯ 45-54 years old
    ◯ 55-64 years old
    ◯ 65-74 years old
    ◯ 75 years or older
    ◯ Prefer not to say

2. What is your gender? *

    *Mark only one oval.*

    ◯ Female
    ◯ Male
    ◯ It's complicated
    ◯ Prefer not to say

3. What is your occupation (job)? *

    _______________________________

4. What is the highest level of education you have received or are pursuing? *

    *Mark only one oval.*

    ◯ Pre-high school
    ◯ High school
    ◯ College
    ◯ PhD (Doctorate)
    ◯ Prefer not to say

5. What is your ethnicity? *

   *Mark only one oval.*

   ○ White
   ○ Black, African American
   ○ Hispanic or Latino
   ○ American Indian or Alaska Native
   ○ Asian Indian
   ○ Chinese
   ○ Filipino
   ○ Japanese
   ○ Korean
   ○ Pacific Islander
   ○ Vietnamese
   ○ Other Asian
   ○ Other

6. What country did you grow up in? *

   ______________________________

7. What country do you live in now? *

   ______________________________

How much do you know about your diet behavior?

8. How much do you agree with the following statements? *

   *Mark only one oval per row.*

   |  | Strongly disagree | Somewhat disagree | Neither agree nor disagree | Somewhat agree | Strongly agree |
   |---|---|---|---|---|---|
   | I eat healthily now | ○ | ○ | ○ | ○ | ○ |
   | I want to eat healthier | ○ | ○ | ○ | ○ | ○ |

9. Please tell us how often you eat the following types of foods. *

   *Mark only one oval per row.*

   |  | Never | Less than once a week | 1-3 times a week | 4-6 times a week | 1 time a day | 2-3 times a day | 4 or more times a day |
   |---|---|---|---|---|---|---|---|
   | Fruits (apples, bananas, oranges, etc.) | ◯ | ◯ | ◯ | ◯ | ◯ | ◯ | ◯ |
   | Vegetables (carrots, mushrooms, etc.) | ◯ | ◯ | ◯ | ◯ | ◯ | ◯ | ◯ |
   | Meat (Chicken, pork, duck, beef, etc.) | ◯ | ◯ | ◯ | ◯ | ◯ | ◯ | ◯ |
   | Fish and Seafood (tuna, shrimp, prawns, etc.) | ◯ | ◯ | ◯ | ◯ | ◯ | ◯ | ◯ |
   | Nuts (almonds, cashews, walnuts, etc.) | ◯ | ◯ | ◯ | ◯ | ◯ | ◯ | ◯ |
   | Dairy (cheese, milk, soy, etc.) | ◯ | ◯ | ◯ | ◯ | ◯ | ◯ | ◯ |
   | Grains (bread, pasta, rice, etc.) | ◯ | ◯ | ◯ | ◯ | ◯ | ◯ | ◯ |
   | Sweets (candy, cookies, cake, etc.) | ◯ | ◯ | ◯ | ◯ | ◯ | ◯ | ◯ |

10. Please tell us how often you eat foods that are cooked in different ways. *

    *Mark only one oval per row.*

    |  | Never | Less than once a week | 1-3 times a week | 4-6 times a week | 1 time a day | 2-3 times a day | 4 or more times a day |
    |---|---|---|---|---|---|---|---|
    | Baked Food | ◯ | ◯ | ◯ | ◯ | ◯ | ◯ | ◯ |
    | Pan Fried Food | ◯ | ◯ | ◯ | ◯ | ◯ | ◯ | ◯ |
    | Deep Fried Food | ◯ | ◯ | ◯ | ◯ | ◯ | ◯ | ◯ |
    | Steamed Food | ◯ | ◯ | ◯ | ◯ | ◯ | ◯ | ◯ |
    | Grilled Food | ◯ | ◯ | ◯ | ◯ | ◯ | ◯ | ◯ |
    | Boiled Food | ◯ | ◯ | ◯ | ◯ | ◯ | ◯ | ◯ |

| How much do you know about nutrition? | Please assume equal portion sizes for below questions. Pay attention to the description of each meal. It is often hard to see all the ingredients in the photograph. The descriptions will help you figure out what the main ingredients of the meal are. |

11. Which meal has more carbohydrates? *

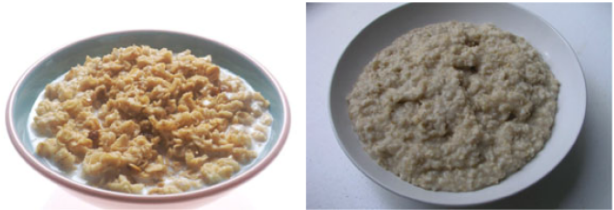

(Left) A single bowl of cooked hot cereal, oatmeal with a cup of whole milk
V.S.
(Right) A single bowl of cooked oatmeal with water

*Mark only one oval per row.*

|   | Definitely Left | Left | Maybe Left | About The Same | Maybe Right | Right | Definitely Right |
|---|---|---|---|---|---|---|---|
| * | ○ | ○ | ○ | ○ | ○ | ○ | ○ |

12. Which meal has more carbohydrates? *

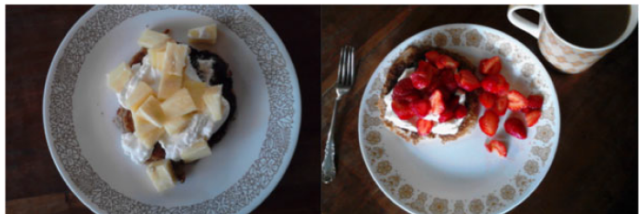

(Left) Wheat flour pancake with greek yogurt and fresh pineapple
V.S.
(Right) Wheat flour pancake with greek yogurt and fresh strawberries

*Mark only one oval per row.*

|   | Definitely Left | Left | Maybe Left | About The Same | Maybe Right | Right | Definitely Right |
|---|---|---|---|---|---|---|---|
| * | ○ | ○ | ○ | ○ | ○ | ○ | ○ |

13. Which meal has more carbohydrates? *

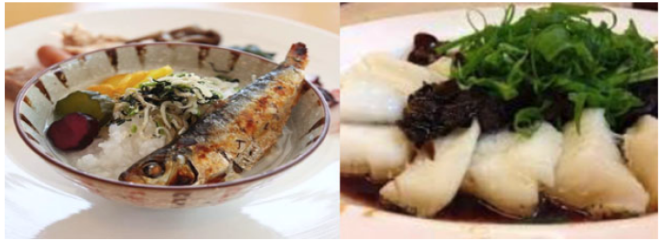

(Left) Porridge with sardines
V.S.
(Right) Fish with mushrooms

*Mark only one oval per row.*

|   | Definitely Left | Left | Maybe Left | About The Same | Maybe Right | Right | Definitely Right |
|---|---|---|---|---|---|---|---|
| * | ◯ | ◯ | ◯ | ◯ | ◯ | ◯ | ◯ |

14. Which meal has more carbohydrates? *

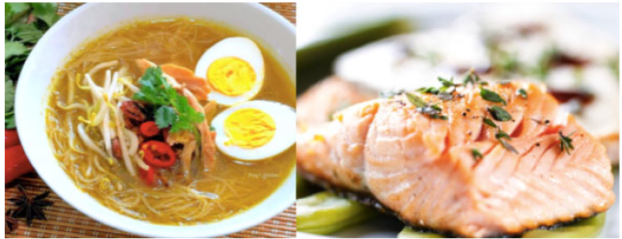

(Left) Mee soto soup with egg and chicken
V.S.
(Right) Salmon

*Mark only one oval per row.*

|   | Definitely Left | Left | Maybe Left | About The Same | Maybe Right | Right | Definitely Right |
|---|---|---|---|---|---|---|---|
| * | ◯ | ◯ | ◯ | ◯ | ◯ | ◯ | ◯ |

15. Which meal has more protein? *

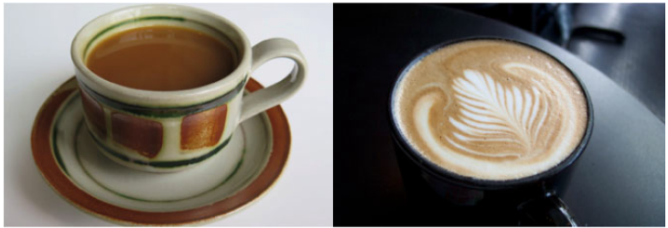

(Left) Cup of coffee with half and half*
\* it is a mixture of cream and milk
V.S.
(Right) Cafe Latte with fat-free milk

*Mark only one oval per row.*

|   | Definitely Left | Left | Maybe Left | About The Same | Maybe Right | Right | Definitely Right |
|---|---|---|---|---|---|---|---|
| * | ◯ | ◯ | ◯ | ◯ | ◯ | ◯ | ◯ |

16. Which meal has more protein? *

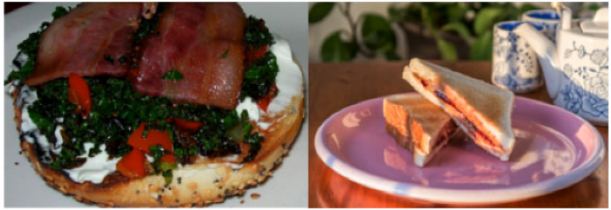

(Left) Kale, red peppers, bacon and cream cheese on a bagel
V.S.
(Right) Peanut butter and jelly sandwich

*Mark only one oval per row.*

|   | Definitely Left | Left | Maybe Left | About The Same | Maybe Right | Right | Definitely Right |
|---|---|---|---|---|---|---|---|
| * | ◯ | ◯ | ◯ | ◯ | ◯ | ◯ | ◯ |

17. Which meal has more protein? *

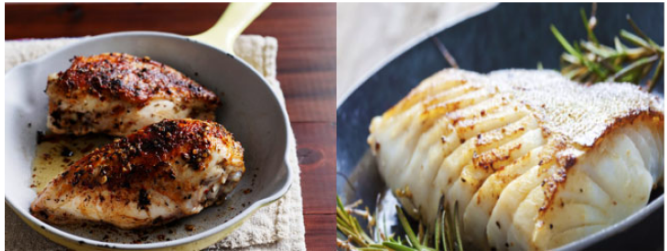

(Left) Grilled chicken breast
V.S.
(Right) Grilled white fish

*Mark only one oval per row.*

|   | Definitely Left | Left | Maybe Left | About The Same | Maybe Right | Right | Definitely Right |
|---|---|---|---|---|---|---|---|
| * | ○ | ○ | ○ | ○ | ○ | ○ | ○ |

18. Which meal has more protein? *

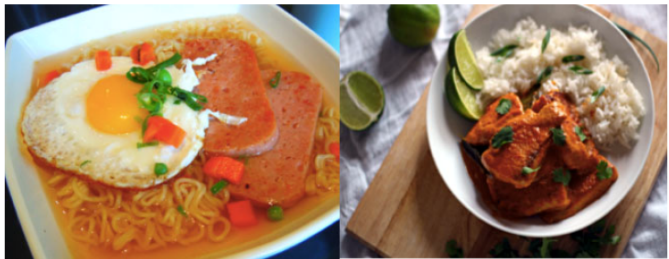

(Left) Noodles with egg and ham
V.S.
(Right) Fish with curry

*Mark only one oval per row.*

|   | Definitely Left | Left | Maybe Left | About The Same | Maybe Right | Right | Definitely Right |
|---|---|---|---|---|---|---|---|
| * | ○ | ○ | ○ | ○ | ○ | ○ | ○ |

19. Which meal has more fat? *

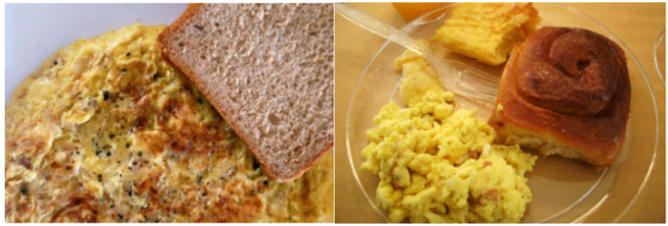

(Left) Masala cheese omlette  
V.S.  
(Right) Scrambled eggs, a bun and orange juice

*Mark only one oval per row.*

|   | Definitely Left | Left | Maybe Left | About The Same | Maybe Right | Right | Definitely Right |
|---|---|---|---|---|---|---|---|
| * | ◯ | ◯ | ◯ | ◯ | ◯ | ◯ | ◯ |

20. Which meal has more fat? *

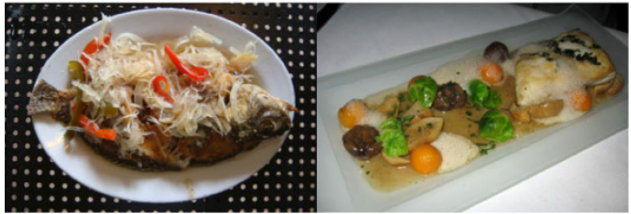

(Left) Roasted tilapia with vegetables  
V.S.  
(Right) Pan roasted halibut* with vegetables  
* it is a flatfish

*Mark only one oval per row.*

|   | Definitely Left | Left | Maybe Left | About The Same | Maybe Right | Right | Definitely Right |
|---|---|---|---|---|---|---|---|
| * | ◯ | ◯ | ◯ | ◯ | ◯ | ◯ | ◯ |

21. Which meal has more fat? *

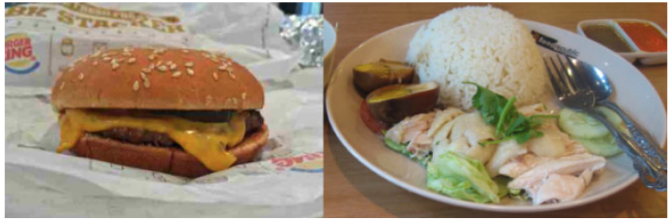

(Left) Cheese burger
V.S.
(Right) Chicken rice

*Mark only one oval per row.*

|  | Definitely Left | Left | Maybe Left | About The Same | Maybe Right | Right | Definitely Right |
|---|---|---|---|---|---|---|---|
| * | ◯ | ◯ | ◯ | ◯ | ◯ | ◯ | ◯ |

22. Which meal has more fat? *

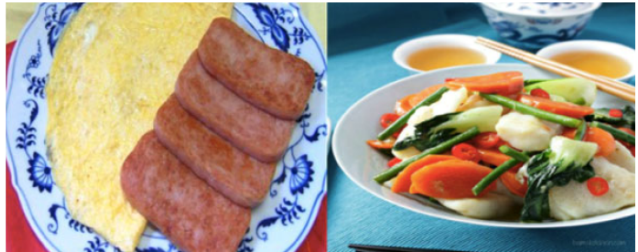

(Left) Egg and spam
V.S.
(Right) White fish with vegetables

*Mark only one oval per row.*

|  | Definitely Left | Left | Maybe Left | About The Same | Maybe Right | Right | Definitely Right |
|---|---|---|---|---|---|---|---|
| * | ◯ | ◯ | ◯ | ◯ | ◯ | ◯ | ◯ |

23. Which meal has more fiber? *

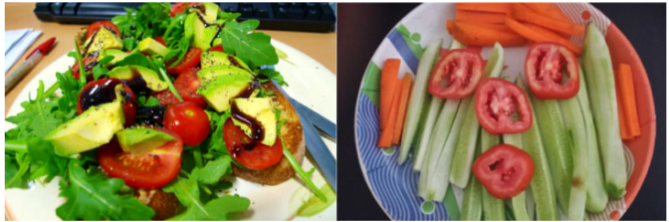

(Left) Mixed salad with arugula, grape tomato, avocado, and balsamic glaze
V.S.
(Right) Carrots and cucumbers

*Mark only one oval per row.*

|   | Definitely Left | Left | Maybe Left | About The Same | Maybe Right | Right | Definitely Right |
|---|---|---|---|---|---|---|---|
| * | ○ | ○ | ○ | ○ | ○ | ○ | ○ |

24. Which meal has more fiber? *

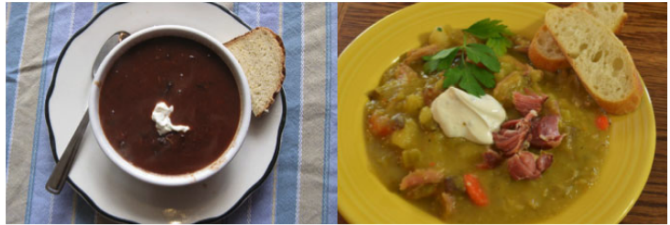

(Left) Black bean soup
V.S.
(Right) Split pea soup

*Mark only one oval per row.*

|   | Definitely Left | Left | Maybe Left | About The Same | Maybe Right | Right | Definitely Right |
|---|---|---|---|---|---|---|---|
| * | ○ | ○ | ○ | ○ | ○ | ○ | ○ |

### A.3 Per-Meal Survey

**For Each Image**
*Required

**Food Reflection Interface [App]**
The App prototype helps you to review and reflect on your eating habits. Please review any information provided and *fill in missing* or *change* details in the following [App] questions as if you are entering information for food log in the User Interface.

[App] Please record the name of this dish. *
Such as: cheeseburger and fries, key lime pie, spaghetti and cheese. NOT lunch, snack, etc.
Sausage and bacon breakfast

[App] Nutrition: Calories (kcal)
Enter a number in kcal if you are confident
600

[App] Nutrition: Carbohydrates (g)
Enter a number in grams if you are confident
50

[App] Nutrition: Proteins (g)
Enter a number in grams if you are confident
20

[App] Nutrition: Fat (g)
Enter a number in grams if you are confident
25

[App] Nutrition: Fiber (g)
Enter a number in grams if you are confident
5

[App] Do you think your meal (or snack) is lower, the same or higher than the nutrient recommendation shown below? *

|  | Average per Meal | |
| --- | --- | --- |
| Macronutrients | Women | Men |
| Calories | 626 | 790 |
| Protein (g) | 35 | 44 |
| Carbohydrates (g) | 70 | 89 |
| Fat (g) | 19 | 24 |
| Fiber (g) | 9 | 10 |

|  | Low | Average | High |
| --- | --- | --- | --- |
| Calorie |  | ● |  |
| Carbohydrates | ● |  |  |
| Protein | ● |  |  |
| Fat |  |  | ● |
| Fiber | ● |  |  |

[App] What food groups does this meal include? *
Please select all that apply.
- [x] Fruits (e.g. oranges, rambutan, bananas)
- [x] Vegetables (e.g. Spinach, peas, broccoli)
- [x] Grains (e.g. Rice, cereal, corn)
- [x] Meat/Fish/Poultry (e.g. shellfish, fish, chicken)
- [ ] Dairy (e.g. Soy, milk, yogurt)

[App] How was this meal cooked? *
Please select all cooking methods that apply.
- [x] Baked
- [x] Pan-fried
- [x] Deep fried
- [ ] Steamed
- [ ] Other:
- [ ] Grilled
- [x] Boiled
- [ ] Roasted
- [ ] None (Raw)

[App] What key ingredients does this meal contain? *
Please list all key raw ingredients. For example, ham, cherry tomato, bread, rice, broccoli, chicken, and more.
Sausage, pineapple, bacon, cinnamon roll

Figure 14: Example per-photo app feedback. Participants were presented with feedback information regarding the meal photo indicated with [App]. In Manual mode, all information is blank or unselected and users have to fill them out (shown here); in Auto mode, all information is pre-filled and users can edit them (shown in Figure 5).

13. From this meal did you learn anything about what you eat? *
Mark only one oval.
1  2  3  4  5
Learned nothing at all  ○ ○ ○ ○ ○  Learned a lot

14. Do you agree or disagree that the information provided is accurate? *
Mark only one oval.
1  2  3  4  5
Strongly Disagree  ○ ○ ○ ○ ○  Strongly Agree

### A.4 Weekly Survey (Reflection)

## Overall Reflection

Now look back at the meals you've eaten and answer the following questions about whether you learned something regarding different aspects of food and your diet. You may refer to meals by their picture number.

* Required

1. 1. Which meal was most unhealthy? *

    Please write the number associated with the image.

    _______________________________

2. Why was it most unhealthy? *

    _______________________________
    _______________________________
    _______________________________
    _______________________________
    _______________________________

3. 2. Which meal was the healthiest? *

    Please write the number associated with the image.

    _______________________________

4. Why was it most healthy? *

    _______________________________
    _______________________________
    _______________________________
    _______________________________
    _______________________________

5. 3. Did you learn anything *interesting* about the meals you ate? *

    Mark only one oval.

    |  | 1 | 2 | 3 | 4 | 5 |  |
    |---|---|---|---|---|---|---|
    | Nothing at all | ○ | ○ | ○ | ○ | ○ | A lot |

6. Why or why not? *

_________________________________________
_________________________________________
_________________________________________
_________________________________________
_________________________________________

**Food Groups**

7. 4. I learned a lot about the *food groups* of the meals that I ate *

   Mark only one oval per row.

   |   | Strongly disagree | Disagree | More or less disagree | Neither disagree nor agree | More or less agree | Agree | Strongly agree |
   |---|---|---|---|---|---|---|---|
   | * | ○ | ○ | ○ | ○ | ○ | ○ | ○ |

8. From which meal did you learn the most about the *food groups*? *

_________________________________________
_________________________________________
_________________________________________
_________________________________________
_________________________________________

9. What did you learn about the *food groups*? *

_________________________________________
_________________________________________
_________________________________________
_________________________________________
_________________________________________

**Cooking Methods**

10. 5. I learned a lot about the *cooking methods* of the meals that I ate *

    Mark only one oval per row.

    |   | Strongly disagree | Disagree | More or less disagree | Neither disagree nor agree | More or less agree | Agree | Strongly agree |
    |---|---|---|---|---|---|---|---|
    | * | ○ | ○ | ○ | ○ | ○ | ○ | ○ |

11. From which meal did you learn the most about the *cooking methods*? *

12. What did you learn about the *cooking methods*? *

### Ingredients

13. 6. I learned a lot about the *ingredients* of the meals that I ate *

    Mark only one oval per row.

    |   | Strongly disagree | Disagree | More or less disagree | Neither disagree nor agree | More or less agree | Agree | Strongly agree |
    |---|---|---|---|---|---|---|---|
    | * | ◯ | ◯ | ◯ | ◯ | ◯ | ◯ | ◯ |

14. From which meal did you learn the most about the *ingredients*? *
    Please write the number associated with the image.

15. What did you learn about the *ingredients*? *

### Macronutrients
Macronutrients include calorie, carbohydrates, protein, fat and fiber.

16. 7. I learned a lot about the *macronutrients* of the meals that I ate *

    Mark only one oval per row.

    |   | Strongly disagree | Disagree | More or less disagree | Neither disagree nor agree | More or less agree | Agree | Strongly agree |
    |---|---|---|---|---|---|---|---|
    | * | ◯ | ◯ | ◯ | ◯ | ◯ | ◯ | ◯ |

17. From which meal did you learn the most about the *macronutrients*? *

    Please write the number associated with the image.

    _______________________________

18. What did you learn about the *macronutrients*? *

    _______________________________
    _______________________________
    _______________________________
    _______________________________
    _______________________________

19. User ID
    PLEASE DO NOT EDIT THE TEXT BELOW

    _______________________________

### A.5 Weekly Survey (Post-Survey)

1. What did you think about using the app feedback? *

   Mark only one oval per row.

   | | Strongly disagree | Disagree | More or less disagree | Neither agree nor disagree | More or less agree | Agree | Strongly agree |
   |---|---|---|---|---|---|---|---|
   | It is easy to use overall | ○ | ○ | ○ | ○ | ○ | ○ | ○ |
   | It is mentally demanding to review each food image | ○ | ○ | ○ | ○ | ○ | ○ | ○ |
   | It is tedious to record/review the name of the dish | ○ | ○ | ○ | ○ | ○ | ○ | ○ |
   | It is tedious to record/review the macronutrients | ○ | ○ | ○ | ○ | ○ | ○ | ○ |
   | It is tedious to record/review the food groups | ○ | ○ | ○ | ○ | ○ | ○ | ○ |
   | It is tedious to record/review the cooking methods | ○ | ○ | ○ | ○ | ○ | ○ | ○ |
   | It is tedious to record/review the key ingredients | ○ | ○ | ○ | ○ | ○ | ○ | ○ |
   | It provides accurate description and details | ○ | ○ | ○ | ○ | ○ | ○ | ○ |
   | It is helpful for understanding about my diet | ○ | ○ | ○ | ○ | ○ | ○ | ○ |
   | I learned a lot about nutrition information | ○ | ○ | ○ | ○ | ○ | ○ | ○ |
   | I learned a lot about my diet behavior | ○ | ○ | ○ | ○ | ○ | ○ | ○ |
   | I learned a lot about how frequently or infrequently I eat certain foods | ○ | ○ | ○ | ○ | ○ | ○ | ○ |
   | I learned a lot about how broad or narrow my diet is overall | ○ | ○ | ○ | ○ | ○ | ○ | ○ |
   | I learned a lot about how I sometimes eat meals that are much healthier or unhealthier than my average diet | ○ | ○ | ○ | ○ | ○ | ○ | ○ |
   | I learned a lot about how healthy or unhealthy my diet is overall | ○ | ○ | ○ | ○ | ○ | ○ | ○ |

2. Please describe any technical problems you had during this study. *

   _______________________________________________
   _______________________________________________

## How much do you know about your diet now?
Now that you have reviewed your diet, please think about your eating behavior again.

3. How much do you agree with the following statements? *

   *Mark only one oval per row.*

   |  | Strongly disagree | Somewhat disagree | Neither agree nor disagree | Somewhat agree | Strongly agree |
   |---|---|---|---|---|---|
   | I eat healthily now | ○ | ○ | ○ | ○ | ○ |
   | I want to eat healthier | ○ | ○ | ○ | ○ | ○ |

4. Please tell us how often you eat the following types of foods. *

   *Mark only one oval per row.*

   |  | Never | Less than once a week | 1-3 times per week | 4-6 times per week | 1 time per day | 2-3 times per day | 4 or more times a day |
   |---|---|---|---|---|---|---|---|
   | Fruits (apples, bananas, oranges, etc.) | ○ | ○ | ○ | ○ | ○ | ○ | ○ |
   | Vegetables (carrots, mushrooms, etc.) | ○ | ○ | ○ | ○ | ○ | ○ | ○ |
   | Meat (Chicken, pork, duck, beef, etc.) | ○ | ○ | ○ | ○ | ○ | ○ | ○ |
   | Fish and Seafood (tuna, shrimp, prawns, etc.) | ○ | ○ | ○ | ○ | ○ | ○ | ○ |
   | Nuts (almonds, cashews, walnuts, etc.) | ○ | ○ | ○ | ○ | ○ | ○ | ○ |
   | Dairy (cheese, milk, soy, etc.) | ○ | ○ | ○ | ○ | ○ | ○ | ○ |
   | Grains (bread, pasta, rice, etc.) | ○ | ○ | ○ | ○ | ○ | ○ | ○ |
   | Sweets (candy, cookies, cake, etc.) | ○ | ○ | ○ | ○ | ○ | ○ | ○ |

5. Please tell us how often you eat foods that are cooked in different ways. *

   *Mark only one oval per row.*

   |  | Never | Less than once a week | 1-3 times per week | 4-6 times per week | 1 time per day | 2-3 times per day | 4 or more times a day |
   |---|---|---|---|---|---|---|---|
   | Baked Food | ○ | ○ | ○ | ○ | ○ | ○ | ○ |
   | Pan Fried Food | ○ | ○ | ○ | ○ | ○ | ○ | ○ |
   | Deep Fried Food | ○ | ○ | ○ | ○ | ○ | ○ | ○ |
   | Steamed Food | ○ | ○ | ○ | ○ | ○ | ○ | ○ |
   | Grilled Food | ○ | ○ | ○ | ○ | ○ | ○ | ○ |
   | Boiled Food | ○ | ○ | ○ | ○ | ○ | ○ | ○ |

## How much do you know about nutrition now?

Please assume equal portion sizes for below questions. Pay attention to the description of each meal. It is often hard to see all the ingredients in the photograph. The descriptions will help you figure out what the main ingredients of the meal are.

6. Which meal has more carbohydrates? *

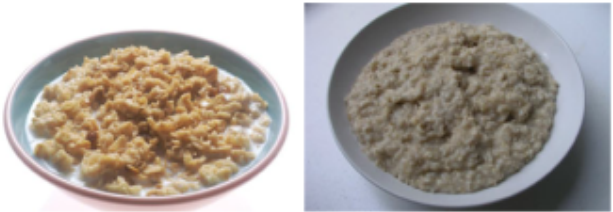

(Left) A single bowl of cooked hot cereal, oatmeal with a cup of whole milk
V.S.
(Right) A single bowl of cooked oatmeal with water

*Mark only one oval per row.*

|   | Definitely Left | Left | Maybe Left | About The Same | Maybe Right | Right | Definitely Right |
|---|---|---|---|---|---|---|---|
| * | ○ | ○ | ○ | ○ | ○ | ○ | ○ |

...

21. Which meal has more fiber? *

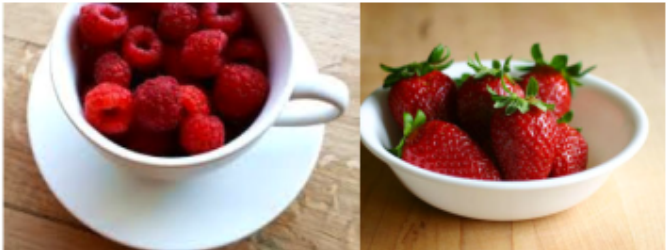

(Left) A cup of raspberries
V.S.
(Right) A cup of strawberries

*Mark only one oval per row.*

|   | Definitely Left | Left | Maybe Left | About The Same | Maybe Right | Right | Definitely Right |
|---|---|---|---|---|---|---|---|
| * | ○ | ○ | ○ | ○ | ○ | ○ | ○ |

22. Have you ever tracked your diet or meals before? If yes, what trackers have you used? *

    Mark only one oval.

    ◯ I have never tracked my diet before
    ◯ Paper diary
    ◯ MyFitnessPal
    ◯ Weight Watchers
    ◯ Other: _______________

| Thank you and Follow-Up | Thank you for completing the pre-survey. Please enter your MTurk worker ID. Note that the same ID is required for the next surveys after you submit your photos. |
|---|---|

23. User ID
    PLEASE DO NOT EDIT THE TEXT BELOW

    _______________

### Photo logging your eating for another week?

If you would like to continue participating in this study for another week, please let us know! You can be rewarded for another *$8.75* to photograph one more week of your meals with *at least 2 meals a day, everyday, for another 7 days*. This will be followed with another post-survey.

If you are interested, please commit to the following statements.

24. I commit to photographing at least *two meals* every day for the next 7 days.
    Type "I commit" if you do.

    _______________

25. I commit to returning after 7 days to complete the follow-up surveys.
    Type "I commit" if you do.

    _______________

Make sure you click the blue "SUBMIT" button to complete the pre-survey!

This content is neither created nor endorsed by Google.

Google Forms

## B  SUPPLEMENTARY RESULTS OF MOBILE FOOD TRACKING INFORMATIVENESS STUDY

We analyzed the background attitudes and logging behavior of participants. These supplement our analysis on user informativeness.

### B.1  Initial Perceptions and Attitudes Towards Healthy Eating, Diet Habits, and Nutrition Knowledge

We analyzed the results from the pre-survey from all 136 respondents. The findings are similar for the subset 53 who continued for the weekly study. About half of participants (56.6%) perceived that they eat healthily, but almost all (91.9%) wanted to eat more healthily (Table 7). Participants mostly frequently ate grains, meat and dairy, but there was a wide distribution of habitual eating of vegetables and fruits (Table 8, left). Participants ate baked foods (i.e., bread) most frequently, followed by grilled, pan fried, and steamed foods, but reported that they rarely ate boiled or deep fried foods (Table 8, right). Since participants were randomly assigned to Feedback Modes, there was no significant difference in pre-survey measures between groups.

We assessed participants' nutrition knowledge to ensure they were able to understand their food and the provided information. We analyzed the participants' performance in the Nutrition Knowledge Test [14] by first binarizing their responses in terms of whether they selected the second (right) item (i.e., rating > 0 or not), then grading whether the selection was correct. This produces a correctness metric that can range between 0 and 1. For the pre-survey, participants demonstrated good understanding (M=0.850, SD=0.304).

Table 7: Participant perception of their eating healthiness (left) and attitude to eat healthier (right).

| I eat healthily now | Count | % | I want to eat healthier | Count | % |
|---|---|---|---|---|---|
| Strong agree | 12 | 8.8% | Strong agree | 71 | 52.2% |
| Somewhat agree | 65 | 47.8% | Somewhat agree | 54 | 39.7% |
| Neither agree nor disagree | 21 | 15.4% | Neither agree nor disagree | 7 | 5.1% |
| Somewhat disagree | 28 | 20.6% | Somewhat disagree | 1 | 0.7% |
| Strong disagree | 10 | 7.4% | Strong disagree | 3 | 2.2% |

### B.2  Per-Meal Food Logs

To help to contextualize participant reflections of meal feedback, we analyzed the meal information logged. Participants with Manual feedback had to write their own information and may have looked-up, estimated, or guessed them, though they would be more able to identify the dishes. Participants with Auto feedback received annotations from the Wizard-of-Oz method, which was rigorously coded from nutrition databases, but may be incorrectly identified. Therefore, both Feedback Modes were prone to some errors in various ways.

We found that the Auto feedback annotated more meals with below average macronutrient levels than Manual feedback (Figure 15). This could be due to participants attempting to eat healthier meals, or eating more snacks rather than full meals; participants with Manual feedback may have over-estimated their macronutrient levels too. Participants ate meals with similar levels of Food Groups across Feedback Mode, with Grains being most common and Fruits being least common (Figure 16, left). The distribution of Cooking Methods was similar between Feedback Mode too, with Baked being most common, followed by Pan-fried, Boiled, Steamed, and Deep fried (Figure 16, middle)). Participants with Manual feedback reported fewer food groups and cooking methods, possibly due to less diligent annotations compared to the researcher annotators. In particular, participants seldom reported bread as being baked. The number of ingredients was the same for both Feedback Modes; participants reported 1 to 27 ingredients per meal (Median=5; Figure 16, right).

Table 8: Participant diet habits regarding eating various food groups (left) and foods cooked with different methods (right).

| Eating Frequency | Food Groups | | | | | | Cooking Methods | | | | | |
|---|---|---|---|---|---|---|---|---|---|---|---|---|
| | Grains | Vegetables | Meat | Seafood | Fruits | Diary | Steamed | Boiled | Baked | Grilled | PanFried | DeepFried |
| ≥4 / day | 4.4% | 0.7% | 2.2% | 0.0% | 0.7% | 2.9% | 0.0% | 0.7% | 0.7% | 0.0% | 0.0% | 0.0% |
| 2-3 / day | 27.9% | 29.4% | 27.9% | 0.7% | 17.6% | 22.1% | 5.1% | 5.1% | 8.1% | 2.2% | 6.6% | 0.7% |
| 1 / day | 22.1% | 22.8% | 23.5% | 5.1% | 20.6% | 28.7% | 6.6% | 5.1% | 14.0% | 8.8% | 10.3% | 3.7% |
| 4-6 / week | 22.8% | 15.4% | 19.9% | 7.4% | 19.9% | 19.1% | 14.7% | 11.0% | 23.5% | 14.7% | 15.4% | 5.1% |
| 1-3 / week | 14.0% | 22.1% | 15.4% | 34.6% | 23.5% | 16.9% | 33.8% | 26.5% | 39.0% | 36.8% | 35.3% | 16.2% |
| <1 / week | 6.6% | 7.4% | 5.1% | 34.6% | 16.9% | 7.4% | 24.3% | 36.0% | 14.0% | 30.9% | 22.8% | 51.5% |
| Never | 2.2% | 2.2% | 5.9% | 17.6% | 0.7% | 2.9% | 15.4% | 15.4% | 0.7% | 6.6% | 9.6% | 22.8% |

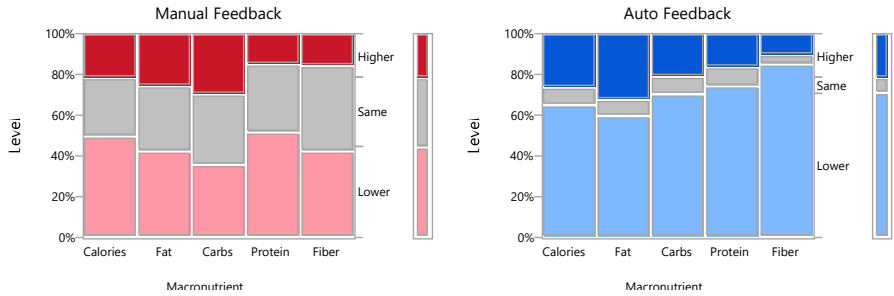

Figure 15: Logged macronutrient levels for participants with Manual (left) and Auto (right) feedback.

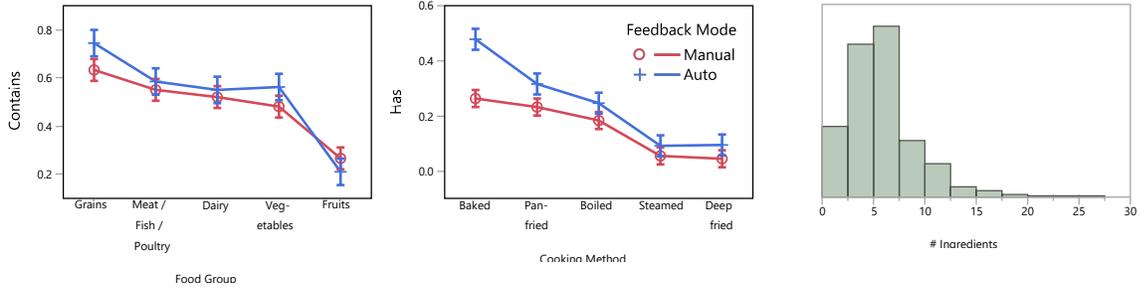

Figure 16: Logged meal information of Food Group (left), Cooking Method (middle) and number of Ingredients (right).

## B.3 Details of Inferential Statistical Models

In our quantitative analyses, we employed the same method for all inferential statistics. We first binarized each measure from bi-polar Likert scale rating to >0 and ≤0, then trained a linear mixed effects model (e.g., Table 4) with Feedback Mode and Week as main fixed effects and Participant as random effect. Additional fixed effects models were included for some analyses.

Table 9: Statistical analysis of perceived accuracy and informativeness ratings regarding per-meal feedback (left) and perceived ease of use, tediousness and informativeness ratings for each week (right). *n.s.* effects are not significant, .01<p<.05 are considered marginally significant.

| Response | Linear Effects Model (Participant random effect) | p>F | $R^2$ |
|---|---|---|---|
| Perceived Accuracy | Feedback Mode + | n.s. | .523 |
|  | Week + | .0011 |  |
|  | Feedback Mode × Week | <.0001 |  |
| Perceived Informativeness | Feedback Mode + | .0387 | .457 |
|  | Week + | <.0001 |  |
|  | Feedback Mode × Week | .0001 |  |

| Response | Linear Effects Model (Participant random effect) | p>F | $R^2$ |
|---|---|---|---|
| Perceived Ease of Use | Feedback Mode + | n.s. | .530 |
|  | Week + | n.s. |  |
|  | Week × Feedback Mode | .0162 |  |
| Perceived Tediousness | Feedback Mode + | n.s. | .537 |
|  | Week + | n.s. |  |
|  | Week × Feedback Mode + | n.s. |  |
|  | Nutrition Information Type + | <.0001 |  |
|  | Info Type × Feedback Mode | n.s. |  |
| Perceived Informativeness | Feedback Mode + | n.s. | .628 |
|  | Information Type + | .0011 |  |
|  | Info Type × Feedback Mode | n.s. |  |
| Informativeness (Nutrition) | Feedback Mode + | n.s. | .567 |
|  | Nutrition Knowledge Info + | <.0001 |  |
|  | Info Type × Feedback Mode | n.s. |  |
| Informativeness (Diet) | Feedback Mode + | n.s. | .537 |
|  | Diet Behavior Information + | <.0001 |  |
|  | Info Type × Feedback Mode | n.s. |  |

Table 10: Statistical analysis of self-reflections with positive and negative valence. *n.s.* effects are not significant, .01<p<.05 are considered marginally significant.

| Response | Linear Effects Model (Participant random effect) | p>F | $R^2$ |
|---|---|---|---|
| Positive Valence Mentioned | Feedback Mode + | .0367 | .313 |
|  | Week + | n.s. |  |
|  | Week × Feedback Mode + | n.s. |  |
|  | Nutrition Knowledge Info + | .0071 |  |
|  | Info Type × Feedback Mode | n.s. |  |

| Response | Linear Effects Model (Participant random effect) | p>F | $R^2$ |
|---|---|---|---|
| Negative Valence Mentioned | Feedback Mode + | n.s. | .219 |
|  | Week + | .0123 |  |
|  | Week × Feedback Mode + | n.s. |  |
|  | Nutrition Knowledge Info + | .0004 |  |
|  | Info Type × Feedback Mode | n.s. |  |

Table 11: Statistical analysis of the number of mentions in self-reflection about contextual information. *n.s.* effects are not significant, .01<p<.05 are considered marginally significant.

| Response | Linear Effects Model (Participant random effect) | p>F | $R^2$ |
|---|---|---|---|
| Context Information Mentioned | Feedback Mode + | <.0001 | .020 |
|  | Week + | n.s. |  |
|  | Week × Feedback Mode + | n.s. |  |
|  | Context Information Type + | n.s. |  |
|  | Context Info × Feedback Mode | .1072 |  |

### B.4 Perceived Burden of Per-Meal Feedback

We combined the ratings for perceived ease of use and mental (non-)demandingness into an overall binarized rating. We analyzed perceived tediousness per nutrition knowledge information type to assess which is most costly and perceived learning with respect to nutrition knowledge and diet behavior information types. As expected, participants perceived Auto feedback as easier to use (Figure 17a) and less tedious (Figure 17b) than Manual feedback, especially after the first week. Reviewing macronutrients was the most tedious, while cooking methods were the least (Figure 17c); there was no difference between Feedback Modes.

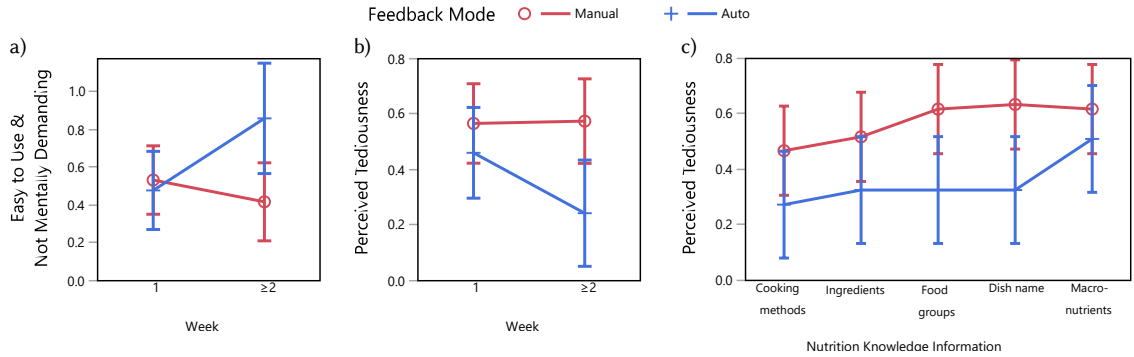

Figure 17: Results of weekly overall perceived tediousness for each Feedback Mode. Response values are binary (0 or 1); error bars indicate 90% confidence interval. Categorical axes sorted in ascending order of y-axis value.

### B.5 Perceived Accuracy of Per-Meal Feedback

Participants reported high perceived accuracy (M=82.4% agreed). Participants with Auto feedback increased their perceived accuracy after the first week, but not those with Manual feedback (Figure 18, left). This suggests their increasing trust in the Auto feedback over time. Participants with Auto feedback had marginally higher perceived informativeness than those with Manual feedback (M=62.7% vs. 42.1%, $p<.0387$).

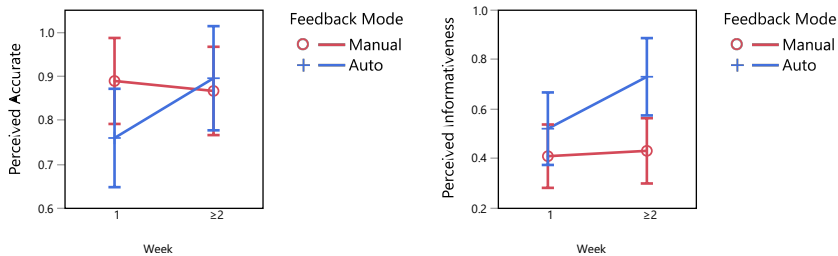

Figure 18: Results of per-meal perceived accuracy (left) and perceived informativeness (right).

### B.6 Perceived Informativeness of Per-Meal Feedback

Not all feedback was perceived as informative (M=46.3% agreed), suggesting the need to not provide feedback all the time. Participants had higher perceived informativeness with Auto feedback after the first week, but not for Manual feedback (Figure 5, right). This suggests that participants could learn more over time with Auto feedback, but not with Manual feedback.

## C EXPLANATION OF INFORMATIVENESS MODEL PREDICTION

### C.1 Model Selected Features

Table 12: Data features used in the best-performing XGBoost machine learning model for predicting perceived learning.

| Feature (Prior Habit) | Definition |
|---|---|
| Prior Eating Habit (Vegetables) | How often to eat vegetables (never, <1/week, 1-3/week, 4-6/week, 1/day, 2-3/day, ≥4/day). |
| Prior Eating Habit (Fruits) | How often to eat fruit. |

| Feature (Current Meal) | Definition |
|---|---|
| Meal Macros (Calorie level) | Calories level in current meal (low, medium, high). |
| Meal Macros (Carbs level) | Carbohydrates level in current meal (low, medium, high). |
| Meal Macros (Protein level) | Protein level in current meal (low, medium, high). |
| Meal Macros (Fat level) | Fat level in current meal (low, medium, high). |
| Meal Macros (Fiber level) | Fiber level in current meal (low, medium, high). |
| Meal Food Group (Grains) | Whether the current meal has grains (none, contains). |
| Meal Food Group (Vegetables) | Whether the current meal has vegetables (none, contains). |
| Meal Food Group (Meat) | Whether the current meal has meat/fish/poultry (none, contains). |
| Meal Food Group (Fruits) | Whether the current meal has fruits (none, contains). |
| Meal Food Group (Diary) | Whether the current meal has diary (none, contains). |
| Meal Food Groups Count | Number of food groups (grains, vegetables, meat/fish/poultry, fruit, dairy) the meal has (1 to 5). |
| Meal Cooking (Baked) | Whether the current meal is or has a part cooked with baking (not, has). |

| Feature (with Previous Meals) | Definition |
|---|---|
| Meal Macros (Calorie level) : Mean[Prev1-Current] | Average meal calorie level in previous 1 and current meals. |
| Meal Macros (Calorie level) : Highest[Prev3-Current] | Highest meal calorie level in previous 3 to current meals. |
| Meal Macros (Protein level) : Highest[Prev3-Current] | Highest meal protein level in previous 3 to current meals. |
| Meal Macros (Fat level) : Highest[Prev3-Current] | Highest meal fat level in previous 3 to current meals. |
| Meal Macros (Calorie level) : Change[Prev1-Current] | Change in calorie level (unchanged, decrease, increase) from previous and current meal. |
| Meal Macros (Fat level) : Change[Prev2-Current] | Change in fat level (unchanged, decrease, increase) from average of previous 2 meals to current meal. |
| Meal Food Group (Vegetables) : Change[Prev2-Current] | Change in presence of vegetables from average of previous 2 meals to current meal. |
| Meal Food Group (Vegetables) : Change[PrevSameMealType-Current] | Change in presence of vegetables (unchanged, decrease, increase) from previous meal of the same type (breakfast, lunch, dinner) to current meal. |
| Meal Ingredients Count : Highest[Prev2-Current] | Highest meal ingredient count in previous 2 to current meals. |
| Meal Cooking (Microwaved) : Mean[Prev1-Current] | Average # meals with microwave cooking in previous and current meals. |
| Meal Cooking (Microwaved) : Mean[Prev3-Current] | Average # meals with microwave cooking in previous 3 to current meals. |
| Meal Cooking (Pan/Air Fried) : Mean[Prev3-Current] | Average # meals with pan/air fried cooking in previous 3 to current meals. |
| Meal Cooking Method (Baked) : SD[Prev2-Current] | Standard deviation of # meals with baked cooking in previous 2 to current meals. |
| Meal Cooking Method (Deep Fried) : SD[Prev2-Current] | Standard deviation of # meals with deep fried cooking in previous 2 to current meals. |
| Meal Cooking (Raw) : SD[Prev3-Current] | Standard deviation of # meals with raw food in previous 3 to current meals. |
| Meal Cooking (Steamed) : Trend[Prev3-Current] | Trend of # meals with steam cooking in previous 3 to current meals. |

## C.2 Instance Explanation of low informativeness example

Table 13 shows the SHAP attribution and Anchor rule explanations of a meal predicted to have low informativeness. Note that even though the meal "Baked salmon, carrots and potatoes" seems healthy, it may not be useful to show feedback. The user is not likely to learn from the meal because: 1) the user's trend of eating steam cooked meals is unchanged, which SalienTrack considers unremarkable; 2) the meal carbohydrate level is higher than the medium level, which is typical for the user; and 3) the recent meals remained similarly not deep fried (low SD), so it is not particularly diverse or interesting.

Table 13: Explanations of a meal of "Baked salmon, carrots and potatoes" predicted to have low informativeness by the model trained on Manual Feedback data. Feature attribution as calculated with SHAP is shown as a tornado plot (vertical bar chart). Solid bars indicate attribution towards informativeness, and hollow bars indicate negative attributions (towards no informativeness). Bars are grey for user profile features, and blue for learning from app feedback. Anchor rules explain with respect to the prediction; in this case, they explain why the meal was predicted to have low informativeness. Only significant rules are shown.

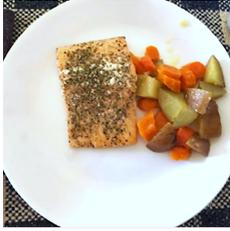

## C.3 Global Explanation

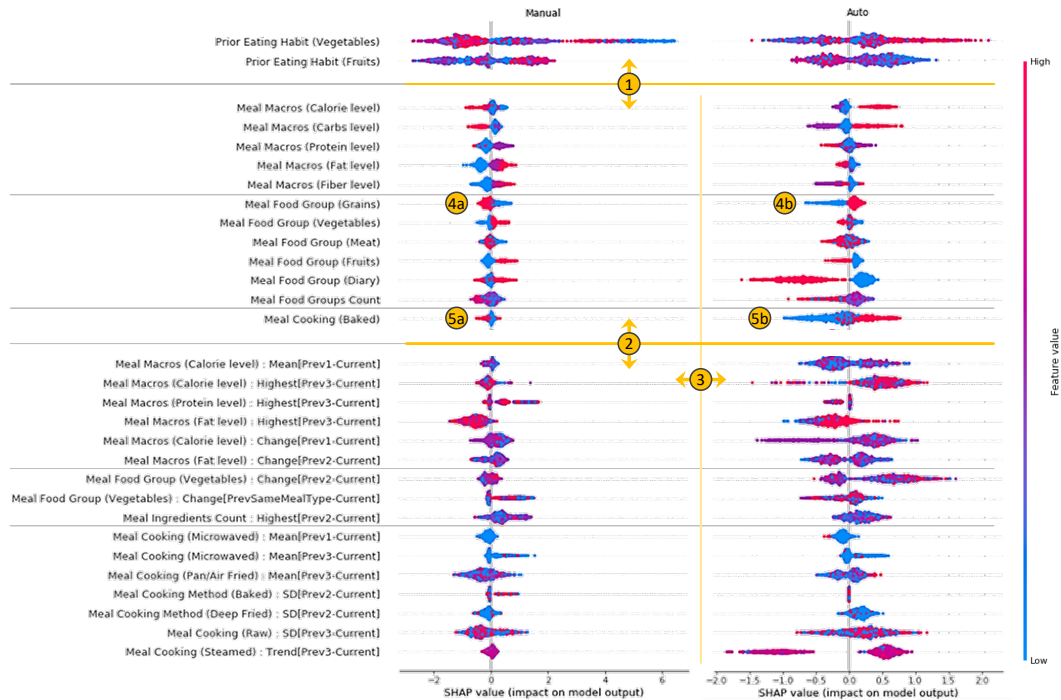

Figure 19: Global influence of features in the informativeness prediction model calculated from SHAP attributions for Manual (left) and Auto (right) feedback models. Each dot presents a test instance. Color represents the relative feature value (red=low, blue=high). Dots to the right of each vertical zero line indicates positive attribution towards high informativeness, and to the left of 0 indicates attribution towards low informativeness. Dots farther from 0 indicate stronger influence for the feature. Numbered yellow annotations indicate insights discussed in the paragraph text.

Aggregating SHAP attributions across a dataset in a scatter plot provides an overview of the model behavior (Figure 19). This global explanation provides insights into which features are the most influential for predicting informativeness. Unlike Lim et al. [55] who used a single Decision Trees for global interpretability, using SHAP globally method is model agnostic, i.e., it can be used to explain any underlying prediction model.

We report key findings annotated in Figure 19. 1) Features about Prior Eating Habits are the most influential (widest spread). For example, users who often eat vegetables tend to learn more from Auto feedback. 2) Features about diet behavior were more influential than features about nutrition knowledge. This agrees with our thematic analysis finding that users learn more about their Diet Behavior than Nutrition Knowledge (Figure 7). 3) Features for the Auto model had higher influence than features for Manual. This suggests that users with Auto feedback were either more surprised about the feedback, or had little need for them (already knew), compared to users with Manual feedback. 4) The direction of informativeness depends on feedback mode: when eating meals with high levels of grains, participants with Manual feedback did not learn much (a), but participants with Auto feedback were surprised (b). 5) The level of influence for each feature varies by feedback mode too: when reviewing whether the current meal had baked cooking, For Manual feedback baked cooking is not influential, but for Auto feedback having baked foods would increase the informativeness of the meal and non-baked foods would be much less informative.

## D TUTORIAL OF EACH FEEDBACK MODE IN FORMATIVE INTERVIEW STUDY OF SALIENTRACK

Table 14: Our introduction of each feedback mode to participants in our interview study.

| Mode | Introduction |
| --- | --- |
| Baseline-Nutrition | Four groups of nutrition information about the current meal are shown, including the levels of calorie and macronutrients, food groups, cooking methods, and ingredients. The levels of calorie and macronutrients are calculated according to the American Dietary Guidelines for adults. Now please read through the information and select the items you find interesting or useful to you. |
| Baseline-Behavior | You can see two groups of information: five items of current meal information, and ten items of diet behavior information, which is accumulated nutrition information of your recent meals, e.g., "high level of calories at most in recent 4 meals". Why these 15 items instead of others? We ran an AI system to analyze the data from our study and found these information items were most relevant to users' perception of learning. In other words, the selected items are the top-15 important ones. Now please read through the information and select the items you find interesting or useful to you. Then I will ask you to compare this mode with the first one. |
| SalienTrack Manual | An AI system selected three items of information for you, because it thought these items would be interesting to you rather than other information for this particular meal. Here it requires you to manually log the selected information items. Note that for different meals, the selected information items may change accordingly. This feature is different from NutritionTrack and DietTrack. Now please tell me the answers to the prompts and whether you find them interesting or useful for you. Then I will ask you to compare this mode with the prior ones. |
| SalienTrack Auto | Another AI system selected three items of information for you with the same logic as the Manual mode. But it shows you the information directly, instead of asking you to manually log. Same as the Manual mode, the selected information items may change for different meals. Now please tell me the answers you would type into the UI and whether you find them interesting or useful for you. Then I will ask you to compare this mode with the prior ones. |
| SalienTrack SemiAuto | The last mode is the hybrid version of the SalienTrack Manual and Auto modes. As you can see, some items are from the Auto mode, while some are from the Manual mode. Why do we combine them? Our prior study showed that both manually logging and automatically receiving information have advantages and disadvantages. For example, manually logging requires more effort from uses, but also makes users think more deeply about their food. You just experienced Manual and Auto modes separately. Now please talk about if this mode makes sense to you or not, and why? |